\documentclass[sigconf,screen]{acmart}

\newif\ifcameraready%

\newif\ifpreprint%
\preprinttrue%

\setcopyright{acmlicensed}
\copyrightyear{2018}
\acmYear{2018}
\acmDOI{XXXXXXX.XXXXXXX}
\acmConference[Conference acronym 'XX]{Make sure to enter the correct
  conference title from your rights confirmation email}{June 03--05,
  2018}{Woodstock, NY}
\acmISBN{978-1-4503-XXXX-X/2018/06}

\usepackage{amsmath}
\usepackage{booktabs}
\usepackage{float}
\usepackage{graphicx}
\usepackage{subcaption}
\usepackage{hyperref}
\usepackage{listings}
\usepackage{mathpartir}
\usepackage{tikz}
\usepackage{url}

\usepackage[capitalise]{cleveref}
\usepackage[dvipsnames,table]{xcolor}

\usepackage{mathtools}
\usepackage{semantic}

\ifpreprint%

\camerareadytrue

\setcopyright{none}
\renewcommand\footnotetextcopyrightpermission[1]{}

\fi

\usepackage{soul}

\ifcameraready%

\newcommand{\changed}[2]{#2}
\newcommand{\removed}[1]{}

\else

\definecolor{darkgreen}{RGB}{0,110,60}

\newcommand{\changed}[2]{\textcolor{red}{\st{#1}} \textcolor{darkgreen}{#2}}
\newcommand{\removed}[1]{\textcolor{red}{\st{#1}}}

\fi

\newcommand*\emptycirc[1][1ex]{%
  \raisebox{-0.2\height}{\tikz\draw (0,0) circle (#1);}}
\newcommand*\halfcirc[1][1ex]{%
  \raisebox{-0.2\height}{%
    \begin{tikzpicture}
      \draw[fill] (0,0)-- (90:#1) arc (90:270:#1) -- cycle ;
      \draw (0,0) circle (#1);
    \end{tikzpicture}}}
\newcommand*\fullcirc[1][1ex]{%
  \raisebox{-0.2\height}{\tikz\fill (0,0) circle (#1);}}

\newcommand{\tightpar}[1]{{\bigskip \noindent\bf #1}}

\newcommand{\modellanguagename}{\textsc{imp}_\textsc{sm}}

\newcommand{\isdefinedas}{\triangleq}
\newcommand{\multiset}[1]{\{\!|#1|\!\}}

\newcommand{\keyword}[1]{\textbf{\texttt{#1}}}

\newcommand{\kwskip}{\keyword{skip}}
\newcommand{\kwalloc}{\keyword{alloc}}
\newcommand{\kwif}{\keyword{if}\;}
\newcommand{\kwelse}{\;\keyword{else}}
\newcommand{\kwwhile}{\keyword{while}\;}
\newcommand{\kwfn}{\keyword{fn}\;}
\newcommand{\kwcall}{\keyword{call}\;}
\newcommand{\kwblock}[1]{\;\{\; #1 \;\}}
\newcommand{\kwfreeze}{\keyword{freeze}\;}

\newcommand{\Store}{\sigma}
\newcommand{\Heap}{H}
\newcommand{\Functions}{F}
\newcommand{\Context}{C}
\newcommand{\State}{S}
\newcommand{\Program}{c}

\newcommand{\Input}{I}
\newcommand{\Output}{Sink}

\newcommand{\sharedmem}{\mathcal{S}}
\newcommand{\trustedmem}{\mathcal{T}}

\newcommand{\nil}{\bot}

\newcommand{\locations}{\mathit{Loc}}
\newcommand{\numbers}{\mathbb{Z}}

\newcommand{\variables}{\mathit{Var}}

\newcommand{\binop}{\oplus}
\newcommand{\step}{\rightarrow}

\newcommand{\observation}{\pi}
\newcommand{\trace}{\tau}

\theoremstyle{definition}
\newtheorem{definition}{Definition}

\lstdefinelanguage{js}{
	keywords={var, if, byte},
	keywordstyle=\color{Maroon}\bfseries,
	ndkeywords={class, export, boolean, throw, implements, this, async, void, new, true, public},
	ndkeywordstyle=\color{darkgray}\bfseries,
	identifierstyle=\color{black},
	sensitive=false,
	comment=[l]{//},
	morecomment=[s]{/*}{*/},
	commentstyle=\color{darkgray}\ttfamily,
	stringstyle=\color{OliveGreen}\ttfamily,
	escapeinside={/*\#}{\#*/},
	morestring=[b]',
	morestring=[b]",
	morestring=[b]`
}
\title{Memory Safety Isn't All You Need: \\
Code Reuse Attacks in Memory Safe Languages}

\ifcameraready%

\author{Eric Cornelissen}
\email{ericco@kth.se}
\affiliation{%
  \institution{KTH Royal Institute of Technology}
  \city{Stockholm}
  \country{Sweden}
}

\author{Musard Balliu}
\email{musard@kth.se}
\affiliation{%
  \institution{KTH Royal Institute of Technology}
  \city{Stockholm}
  \country{Sweden}
}

\else

\author{Anonymous author(s)}
\email{anon@ymo.us}
\affiliation{%
  \institution{Anonymous Institution}
  \city{Anonymous City}
  \country{Anonymous Country}
}

\fi

\begin{document}

\begin{abstract}
Memory safety rules out classical code reuse attacks that rely on arbitrary memory corruption, but does not eliminate code reuse altogether. Attacks such as object injection, prototype pollution, class pollution, DOM clobbering, and DOM script gadgets arise in memory-safe languages without violating the underlying memory safety abstraction.

We show that these seemingly different attacks share a common structure: Attacker-controlled data influences accesses to shared runtime state, such as runtime type information, prototype or class state, and the DOM, and a value fetched from that state subsequently influences a security-sensitive operation. Drawing on this insight, we develop a language-agnostic framework that decomposes an attack into a weakness, ending in a fetch from shared runtime state, and a gadget, which propagates the fetched value to a sensitive sink. We show that the framework captures five established attack classes and use it to derive stateless, mutating, and recycling weaknesses, together with simple, vertical, and horizontal gadget structures. We further systematize existing defenses into three common strategies---mediation, minimization, and constraint---and use them to derive candidate defense directions. Finally, we develop a model language that gives a formal account of code-reuse weaknesses and potential conditions for gadgets and end-to-end vulnerabilities.
\end{abstract}

\ccsdesc[300]{Security and privacy~Web application security}
\ccsdesc[300]{Security and privacy~Information flow control}

\keywords{Code reuse attacks, Security condition, Information flow control, Web application security}

\maketitle

\ifpreprint%

\pagestyle{plain}

\fi

\section{Introduction}\label{section:introduction}
There is a long history of code-reuse attacks in memory unsafe languages~\cite{crane2018continuing}. Starting with return-to-libc, attackers reused existing privileged functions when direct code injection became infeasible. This evolved into Return-Oriented Programming (ROP) and Jump-Oriented Programming (JOP), where existing code fragments, or \emph{gadgets}, are chained to realize attacker-chosen behavior.

\begin{figure}[t]
  \centering
  \includegraphics[width=0.65\columnwidth]{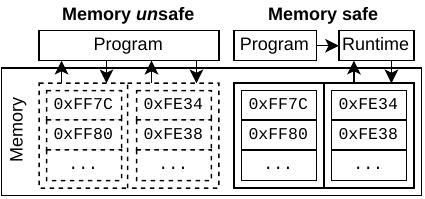}
  \caption{
    Memory access with or without memory safety.
  }\label{figure:unsafe-safe}
\end{figure}

Memory safety rules out classical code reuse mechanisms that rely on arbitrary memory corruption. \Cref{figure:unsafe-safe} illustrates this distinction. In the memory-unsafe setting (left), the program may access memory outside its intended abstraction, enabling arbitrary reads and writes. In the memory-safe setting (right), accesses are restricted to valid objects and locations, preventing such attacks.

Yet code-reuse attacks remain possible in memory safe languages~\cite{dahse2014code,chen2024efficient,shcherbakov2021serialdetector,khaled2023prototype}. Consider the locations \texttt{0xFE34} and \texttt{0xFE38} in \Cref{figure:unsafe-safe}. Memory safety prevents an access to one location from escaping its bounds and reaching the other location, but it does not necessarily prevent attacker-controlled input from determining which valid location the program accesses. Language-level mechanisms such as dynamic dispatch, dynamic property lookup, reflection, and deserialization can expose precisely such choices. An attacker may therefore influence where the program reads from or writes to, as well as the value that is written, while remaining within the memory safety abstraction.

The relevant state in these attacks is the \emph{shared runtime state} which is made accessible across otherwise distinct program components through runtime abstractions, such as runtime type information, prototype objects, class and module state, and the DOM. Object injection~\cite{dahse2014code,chen2024efficient,shcherbakov2021serialdetector}, for example, allows attacker-controlled data to influence which existing type and behavior is selected during dynamic dispatch. Prototype~\cite{arteau2018prototype} and class pollution~\cite{khaled2023prototype} instead allow attacker-controlled accesses to navigate to shared state, modify it, and affect later accesses.

\changed{Across these attacks, we observe a common structure.}{These attacks exhibit a common structure.} A \emph{weakness} captures attacker influence over interactions with shared runtime state and ends in a fetch. The fetched value then serves as the interface to a \emph{gadget}, namely pre-existing program behavior that propagates it through data, control, or call flow to a security sensitive operation. In this way, the attacker reuses existing program behavior without supplying executable code.

Prior work has primarily studied such attacks individually, producing techniques for their detection, exploitation, and mitigation. What remains missing is a language-agnostic account of their common structure that can support reasoning about attacks and defenses across languages. We address this gap by constructing a framework from five established attack classes, using it to derive additional attack structures and defense directions, and developing a model language that gives a semantic account of code reuse under memory safety. In sum, our contributions are:

\begin{itemize}
  \item We develop a language-agnostic framework for code reuse attacks under memory safety and demonstrate its applicability to five established attack classes. We further illustrate its use for analyzing language mechanisms through a Go case study (\Cref{section:description}).
  \item We use the framework to derive stateless, mutating, and recycling weaknesses and simple, vertical, and horizontal gadget structures, exposing attack patterns beyond the established classes (\Cref{section:categorization}).
  \item We map existing defenses to the framework and identify three common strategies, namely mediation, minimization, and constraint, which we use to derive candidate defense directions (\Cref{section:defenses}).
  \item We develop $\modellanguagename$, a model language with explicit shared runtime state, and give a formal characterization of code reuse weaknesses together with potential conditions for gadgets and end-to-end vulnerabilities. We validate the model against the five established attack classes and provide an executable interpreter implementing its syntax and operational semantics (\Cref{section:model}).
\end{itemize}

\section{Code Reuse Attack Framework}\label{section:description}
This section develops a language-agnostic framework to describe code reuse attacks in memory safe languages.
Rather than characterizing attacks by language-specific mechanisms such as prototypes, reflection, or deserialization, the framework abstracts them in terms of how attacker-controlled data influences shared runtime state and how values obtained from that state subsequently affect program execution.
We derive the framework from common mechanisms exhibited by established attacks, including \emph{object injection} in PHP~\cite{dahse2014code,shahriar2016object}, Java~\cite{chen2024efficient,luo2024rev}, and C\#~\cite{shcherbakov2021serialdetector}, \emph{prototype pollution} in JavaScript~\cite{arteau2018prototype,shcherbakov2023silent,liu2024undefined}, \emph{class pollution} in Python~\cite{khaled2023prototype,liu2026first}, and \emph{DOM clobbering}~\cite{khodayari2023dom,liu2025domino} and \emph{DOM script gadgets}~\cite{lekies2017code,drescher2025dom} on the web.

We first define the primitives and attack structure captured by the framework. We then show that five established attack families can be expressed as paths through it, illustrating that seemingly different vulnerabilities share a common underlying structure. Finally, we demonstrate how the framework can guide the analysis of a programming language by applying it to Go.

\subsection{Understanding Code Reuse}\label{section:description:definition}
Memory safety prevents programs from interpreting arbitrary values as memory references and constrains accesses to valid objects and locations. It does not, however, prevent untrusted data from influencing which valid location is accessed, which valid location is modified, or which value is stored there. Language features such as dynamic property lookup, reflection, dynamic dispatch, and deserialization intentionally expose such choices at the abstraction level. Our key insight is that code-reuse attacks in memory-safe languages exploit language-level mechanisms for selecting and manipulating valid program state rather than violating the underlying memory-safety abstraction.

Based on this insight, a memory-safe code reuse attack can manifest when untrusted data is used to determine the access location, store location, or both during program execution. The value in memory may, but is not required to, be controlled by the attacker. Importantly, the memory in question needs to be \emph{shared}. The relevant shared memory is shared runtime state: state reachable through runtime-provided abstractions and potentially accessed by program components that do not otherwise communicate directly. Examples include prototype objects, class and module state, runtime type metadata, and the DOM. We remark that shared memory does not refer to thread- or process-level shared memory; it denotes state made available across otherwise distinct parts of a program through a common runtime abstraction. We formalize this notion in \Cref{section:model} using reachability from runtime-provided roots.

Attacker influence over shared state is not sufficient by itself to constitute a code-reuse attack. The relevant shared-state interaction must causally influence a security-sensitive operation. Specifically, a value obtained from shared state must affect either the data consumed by the sensitive operation, whether the operation executes, or which pre-existing function is invoked on the path to it. This causal connection separates a code-reuse vulnerability from an unrelated shared-state weakness occurring in the same program.

Note that, as with classical code reuse attacks, the attacker is limited to inputting data (raw bytes or primitive data) into the program. Direct execution of attacker-controlled code is not considered a code reuse attack.

\begin{figure}[t]
  \centering
  \includegraphics[width=0.8\columnwidth]{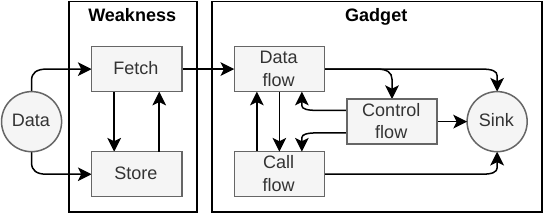}
  \caption{
    Framework for code-reuse attacks. Attacks begin with attacker-controlled data and end at a security-sensitive sink. The weakness consists of attacker-influenced Fetch and tore operations on shared runtime state and terminates in a Fetch; the gadget propagates the fetched value to the sink.
  }\label{figure:framework}
\end{figure}

\begin{table}[]
  \begin{tabular}{ l l }
      \multicolumn{2}{c}{\underline{\textit{Weakness}}} \\
    $\text{Data}\to\text{Fetch}$          & $\texttt{fetch(d)}$                                      \\
    $\text{Data}\to\text{Store}$          & $\texttt{store(d, d')}$                                  \\
    $\text{Store}\to\text{Fetch}$         & $[\texttt{store(d, d')} \Rightarrow] \texttt{ fetch(d)}$ \\
    $\text{Fetch}\to\text{Store}$         & $\texttt{store(d, fetch(d'))}$                           \\
      \midrule
    $\text{Fetch}\to\text{Data flow}$     & $\texttt{x}[\texttt{=fetch(d)}]$                         \\
    $\text{Data flow}\to\text{Sink}$      & $\texttt{sink(x)}$                                       \\
    $\text{Data flow}\to\text{Ctrl flow}$ & $\texttt{if (x) [...]}$                                  \\
    $\text{Data flow}\to\text{Call flow}$ & $\texttt{x()}$                                           \\
    $\text{Ctrl flow}\to\text{Sink}$      & $[\texttt{if (x)}] \texttt{ sink()}$                     \\
    $\text{Ctrl flow}\to\text{Data flow}$ & $[\texttt{if (x)}] \texttt{ x=y}$                        \\
    $\text{Ctrl flow}\to\text{Call flow}$ & $[\texttt{if (x)}] \texttt{ f()}$                        \\
    $\text{Call flow}\to\text{Data flow}$ & $\texttt{x=f()}$                                         \\
    $\text{Call flow}\to\text{Sink}$      & $\texttt{x()} \Rightarrow \texttt{sink()}$               \\
      \multicolumn{2}{c}{$\overline{\textit{Gadget}}$} \\
  \end{tabular}
\end{table}

\subsection{Framework}\label{section:description:framework}
\Cref{figure:framework} models a code-reuse attack as a path from attacker-controlled data to a security-sensitive sink. The graph distinguishes interactions with shared runtime state from the ways in which their effects propagate through program execution. \emph{Fetch} or \emph{Store} represent reads from and writes to shared runtime state, while \emph{Data flow}, \emph{Control flow}, and \emph{Call flow} represent three ways in which values obtained from that state can influence subsequent execution.

We divide a code-reuse attack into two components: a \emph{weakness} and a \emph{gadget}, following terminology from classical code-reuse attacks~\cite{crane2018continuing}. The weakness captures how attacker-controlled data influences interactions with shared runtime state. It begins at \emph{Data} and terminates in a \emph{Fetch} from shared runtime state; one or more Store or Fetch operations may precede the final Fetch. The value produced by the final Fetch is the interface between the weakness and the gadget. A weakness alone does not constitute an attack.

The gadget consists of pre-existing program behavior that propagates the value exposed by the final Fetch to a security-sensitive operation. The fetched value may directly be consumed by the operation (\emph{Data flow}), determine whether particular code executes (\emph{Control flow}), or influence which pre-existing function is invoked (\emph{Call flow}). These forms of influence may be combined, e.g., a call-flow change may occur indirectly through a control flow change, but must ultimately reach a security-sensitive operation, represented by \emph{Sink}. Merely having a weakness and a sensitive operation in the same program is insufficient: the value exposed by the former must influence whether or how the latter is performed.

The Fetch and Store primitives are semantic abstractions rather than specific program constructs. They cover explicit reads and writes as well as equivalent interactions performed implicitly by a language or runtime, for example through property lookup, dynamic dispatch, reflection, or other language-level mechanisms. This abstraction allows the same framework path to capture attacks that concretely differ substantially across languages.

Each edge in \Cref{figure:framework} represents a minimal relationship between two framework primitives. The code patterns shown below the figure illustrate representative implementations of these relationships rather than complete attack programs. For example, the edge $\texttt{Control flow}\to\texttt{Call flow}$ corresponds to a branching point leading to a function call. The metavariable \texttt{d} denotes potentially attacker-controlled data, while \texttt{x} denotes a value whose influence is propagated by the attack. The abstract operations \texttt{fetch}, \texttt{store}, and \texttt{sink}, denote, respectively, accessing shared runtime state, modifying shared runtime state, and performing a security-sensitive operation. Additional statements or equivalent language constructs between the illustrated operations are not precluded.

\begin{figure}[]
\centering
\begin{lstlisting}[caption={A minimal object injection attack (Java). \texttt{Cmd} is an interface with a method \texttt{exec}.},label={listing:object-injection}]
byte[] i1 = new byte[]{/*PAYLOAD*/};
Cmd c = (Cmd) deserialize(i1);
Runnable x = c::exec;
x.run();
class Bkp implements Cmd {
  public void exec() { DB.Backup(); } }
class OSCmd implements Cmd { String[] cmd;
  public void exec() {
    new ProcessBuilder(this.cmd).start(); } }
\end{lstlisting}
\end{figure}

\subsection{Revisiting Known Attacks}\label{section:description:known-attacks}
We now assess the expressiveness of the framework by applying it to five established attack classes across several programming languages and environments. For each class, we identify the shared runtime state involved, the attacker-controlled operation pertaining to the weakness, and the path through the gadget that reaches the sensitive operation. The purpose is not to reproduce the full mechanics of each attack, but to show that their language-specific mechanisms instantiate the same small set of framework primitives.

\subsubsection{Object Injection}\label{section:description:object-injection}
An object-injection vulnerability arises when attacker-controlled (serialized) data causes a program to reconstruct an unintended object type whose pre-existing security-sensitive behavior is subsequently invoked~\cite{esser2010utilizing}. Such vulnerabilities affect languages and runtimes that support reconstructing objects from serialized representations, including PHP, Java, and C\#~\cite{dahse2014code,shcherbakov2021serialdetector}. The relevant shared runtime state consists of the types (classes) available to the running program and their associated method implementations.

In terms of our framework, object injection exhibits a \emph{stateless} weakness. Attacker-controlled input influences which existing type, and consequently which behavior, is selected during object reconstruction and method resolution. This corresponds to the path $\texttt{Data}\to\texttt{Fetch}$. Here, Fetch abstracts the runtime selection of behavior associated with the attacker-chosen object type, rather than an explicit program-level read of a method pointer.

\Cref{listing:object-injection} shows a snippet of an object injection vulnerability (based on Listings 1 and 2 of~\cite{shcherbakov2021serialdetector}).
The intended behavior is to instantiates a benign instance of \texttt{Cmd} (line 2), such as \texttt{Bkp}, and execute it (line 3--4).
When attacked, the following path through the framework graph is traced
\begin{equation*}
  \begin{aligned}
    \text{Data} & \xrightarrow{\texttt{c::exec}} \text{Fetch}
                  \xrightarrow{\texttt{Runnable x = } \cdots} \text{Data flow} \\
                & \xrightarrow{\texttt{x.run()}} \text{Call flow}
                  \xrightarrow{\texttt{new ProcessBuilder(this.cmd)}} \text{Sink}
  \end{aligned}
\end{equation*}
In particular, an attacker constructs a byte stream representing an object (line 1), in this case of type \texttt{Cmd}, allowing it to pick the reference to a method the program will invoke, in this case \texttt{c::exec} (line 3--4). By encoding an \texttt{OSCmd} object, the attacker can cause a new process to be started (line 9).

The Fetch therefore captures the selection of pre-existing behavior determined by the attacker-selected receiver, while the resulting callable behavior forms the interface between the weakness and the gadget. The gadget then reuses the program's existing invocation mechanism and the pre-existing \texttt{OSCmd.exec} implementation to reach the sink; no attacker-supplied executable code is required.

\begin{figure}[]
\centering
\begin{lstlisting}[caption={A minimal prototype pollution attack (JavaScript). Both \texttt{obj1} and \texttt{obj2} are plain JavaScript objects.},label={listing:prototype-pollution}]
var i1 = "__proto__", i2 = "p", i3 = PAYLOAD
obj1[i1][i2] = i3
var x = obj2.p || "..."
eval(x)
\end{lstlisting}
\end{figure}

\subsubsection{Prototype Pollution}\label{section:description:prototype-pollution}
A prototype pollution vulnerability is a code reuse attack affecting prototype-based languages, most notably JavaScript~\cite{arteau2018prototype}. Unlike object injection, it relies on a \emph{stateful} weakness: attacker-controlled input first modifies shared runtime state, and the modified state is subsequently fetched by the program. In JavaScript, prototype objects provide such shared state because properties defined on a prototype can be observed through property lookups on objects that inherit from it.

The weakness arises when attacker-controlled property names allow the program to navigate to a prototype and modify one of its properties. A later property lookup on another object fetches the modified value through the prototype chain. Thus, the weakness follows the path $\texttt{Data}\to\texttt{Store}\to\texttt{Fetch}$, with the final Fetch exposing the polluted value to the gadget.

\Cref{listing:prototype-pollution} illustrates this pattern (based on Listing 1 and 2 of~\cite{shcherbakov2023silent}).
The intended behavior of line 2 is to update a nested property of an object, e.g., a user's profile as \texttt{obj1["profile"]["name"] = "Bob"}, and then dynamically evaluate a trusted script (line 3--4).
When attacked, the following path through the graph is traced
\begin{equation*}
 \begin{aligned}
    \text{Data} & \xrightarrow{\texttt{obj1[i1][i2] = i3}} \text{Store}
                  \xrightarrow{\texttt{obj2.p}} \text{Fetch} \\
                & \xrightarrow{\texttt{var x = } \cdots} \text{Data flow}
                  \xrightarrow{\texttt{eval(x)}} \text{Sink}
  \end{aligned}
\end{equation*}
In particular, an attacker must obtain a reference to the prototype, in this case through the \texttt{\_\_proto\_\_} property, and store something, in this case the attacker's payload at the attacker-controlled property \texttt{p} (line 2). The data must be fetched by the program, in this case through an explicit access of the inherited property \texttt{p} (line 3), and used in some insecure way, in this case as an argument to the \texttt{eval} sink (line 4). Here, the Store captures the attacker-induced modification of shared runtime state, while \texttt{obj2.p} is the final Fetch of the weakness and forms the interface to the gadget.

\begin{figure}[]
\centering
\begin{lstlisting}[caption={A minimal class pollution attack (Python). The program defines a \texttt{class A} and \texttt{class B} and \texttt{o} is an \texttt{A}.},label={listing:class-pollution}]
i1, i2, i3, i4, i5, i6="__class__", "__init__", "__globals__", "B", "p", PAYLOAD
setattr(getattr(getattr(getattr(o, i1), i2), i3)[i4], i5, i6)
x = B.p
eval(x)
\end{lstlisting}
\end{figure}

\subsubsection{Class Pollution}\label{section:description:class-pollution}
A class pollution vulnerability is a code reuse attack similar to prototype pollution but instead affects the Python programming language~\cite{khaled2023prototype}.
Python shares an interconnected object graph that includes instances, class objects, function objects, and module/global namespaces. Metaprogramming APIs such as \texttt{getattr} and \texttt{setattr} can use runtime strings to navigate and modify this graph.

A class pollution weakness arises when attacker-controlled data enable navigation from an initially available object to shared class or module state and subsequent mutation of that state. A gadget later reads the affected field or variable through ordinary program behavior and uses it at a sensitive operation.

\Cref{listing:class-pollution} shows a snippet of a class pollution vulnerability (based on Section 3.1 of~\cite{liu2026first}).
The intended behavior is similar to that of the prototype pollution example.
When attacked, the following path through the framework graph is traced
\begin{equation*}
  \begin{aligned}
    \text{Data} & \xrightarrow{\texttt{setattr(getattr(}\cdots\texttt{)[i4], i5, i6)}} \text{Store}
                  \xrightarrow{\texttt{B.p}} \text{Fetch} \\
                & \xrightarrow{\texttt{x = } \cdots} \text{Data flow}
                  \xrightarrow{\texttt{eval(x)}} \text{Sink}
  \end{aligned}
\end{equation*}
In particular, an attacker must obtain a reference to a \changed{dunder property\footnote{The conventional name for meta properties starting and ending with two `\_'.}}{special attribute \cite{pythonlanguagereference} (a.k.a. dunder property)}, in this case \texttt{\_\_init\_\_} and \texttt{\_\_globals\_\_}, and store something, in this case the attacker's payload using the attacker-controlled name \texttt{p} (line 2). The data must be fetched by the program, in this case as a property of the class \texttt{B} (line 3), and used in some insecure way, in this case as an argument to \texttt{eval} (line 4).

\subsubsection{DOM Clobbering}\label{section:description:dom-clobbering}
A DOM clobbering vulnerability is a type of code reuse attack where a web script accesses properties of the \emph{shared} Document Object Model (DOM) that contains (sanitized) user generated content through so-called \emph{named properties}~\cite{khodayari2023dom}. These properties are created by the browser on certain DOM components for elements in the DOM with a \texttt{name} or \texttt{id} property.

The weakness is caused by the browser changing the contents of, e.g., the \texttt{window} or \texttt{document} object based on the contents of the DOM, in some cases shadowing existing properties.
A gadget is a client-side script that inadvertently accesses a named property when performing a sensitive operation.

\Cref{listing:dom-clobbering} shows a snippet of a DOM clobbering vulnerability (based on CVE-2024-43788~\cite{liu2025domino}).
The intended behavior is to insert a DOM element with user content (line 2--4) and then load a new script from the same source as the current script (line 5--8).
When attacked, the following path through the framework graph is traced
\begin{equation*}
  \begin{aligned}
    \text{Data} & \xrightarrow{\texttt{document.body.appendChild(el)}} \text{Store}
                  \xrightarrow{\texttt{document.currentScript}} \text{Fetch} \\
                & \xrightarrow{\texttt{var x = } \cdots} \text{Data flow}
                  \xrightarrow{\texttt{script.src = x.src}} \text{Sink}
  \end{aligned}
\end{equation*}
In particular, an attacker must control part of the DOM, in this case an \texttt{<img>} (line 2-4). This is fetched by the program through a named property access, in this case \texttt{currentScript} (line 5), and used for a security sensitive operation, in this case to load a new web script (line 7).

\begin{figure}
\centering
\begin{lstlisting}[caption={A minimal DOM clobbering attack (JavaScript).},label={listing:dom-clobbering}]
var i1 = "name", i2 = "currentScript", i3 = "src",  i4 = PAYLOAD
var el = document.createElement("img")
el.setAttribute(i1,i2); el.setAttribute(i3,i4)
document.body.appendChild(el)
var x = document.currentScript
var script = document.createElement("script")
script.src = x.src
document.head.appendChild(script)
\end{lstlisting}
\end{figure}

\subsubsection{DOM Script Gadgets}\label{section:description:dom-script-gadgets}
A DOM script gadget is a type of code reuse attack where a web script reads data from a DOM containing user generated content and uses it in a sensitive operation~\cite{lekies2017code}.

DOM script gadgets differ from DOM clobbering in where the unexpected interaction occurs. The weakness does not rely on named-property resolution. Instead, attacker-controlled content is intentionally stored in the DOM, where the insertion is safe in isolation. A later script treats the DOM as trusted shared state, fetches the content, and propagates it to a sensitive operation. Thus, both attacks use the DOM as shared state, but DOM clobbering abuses implicit access to the DOM, whereas DOM script gadgets abuse how explicitly stored DOM data is later interpreted.

\Cref{listing:dom-script-gadgets} shows a snippet of a DOM script gadget vulnerability (based on Listing 18 of~\cite{lekies2017code}).
The intended behavior is to set the text of a button based on an attribute (line 5--6) after having inserted a DOM element with user content (line 2--4).
When attacked, this traces the following path through the framework graph
\begin{equation*}
  \begin{aligned}
    \text{Data} & \xrightarrow{\cdots \texttt{.appendChild(el)}} \text{Store}
    \xrightarrow{\cdots \texttt{.getElementById("button")}} \text{Fetch} \\
                & \xrightarrow{\texttt{var x = } \cdots} \text{Data flow}
                  \xrightarrow{\texttt{x.innerHTML = x.getAttribute(} \cdots \texttt{)}} \text{Sink}
  \end{aligned}
\end{equation*}
In particular, an attacker must control part of the DOM, in this case a \texttt{<div>} (line 2--4). This data is fetched by the program using standard DOM APIs, in this case \texttt{getElementById} (line 5), and used for a security sensitive operation, in this case as an assignment to the \texttt{Element.innerHTML} sink (line 6).

\subsubsection{Framework Results}\label{section:description:conclusion}
Applying the framework to the five established attack classes reveals a common structure despite substantial differences in their language-level mechanisms. In each case, attacker-controlled data influences an interaction with shared runtime state, and the weakness terminates in a Fetch that exposes a value to a gadget. The gadget then propagates this value through data, control, or call flow to a security-sensitive sink. This decomposition provides a common interface, i.e., the value produced by the final Fetch, between the weakness and the gadget.

The framework also exposes two dimensions along which the weaknesses differ. First, they can be \emph{stateless} or \emph{stateful}.
Object injection directly influences a Fetch, following $\texttt{Data}\rightarrow\texttt{Fetch}\rightarrow\cdots\rightarrow\texttt{Sink}$, whereas prototype pollution, class pollution, DOM clobbering, and DOM script gadgets first modify shared runtime state and later fetch from it, following $\texttt{Data}\rightarrow\texttt{Store}\rightarrow\cdots\rightarrow\texttt{Sink}$. Second, attacks differ in what attacker-controlled data influences: the location from which a value is fetched, the location to which a value is stored, or the value being stored. The five attacks instantiate only some combinations of these possibilities. In \Cref{section:categorization}, we systematically explore this design space and show that the framework yields additional weakness and gadget structures beyond those exhibited by the known attack classes.

\begin{figure}[]
\centering
\begin{lstlisting}[caption={A minimal DOM script gadget attack (JavaScript).},label={listing:dom-script-gadgets}]
var i1 = "id", i2 = "button", i3 = "data-text", i4 = PAYLOAD
var el = document.createElement("div")
el.setAttribute(i1,i2); el.setAttribute(i3,i4)
document.body.appendChild(el)
var x = document.getElementById("button")
if (x) x.innerHTML = x.getAttribute("data-text")
\end{lstlisting}
\end{figure}

\subsection{Case Study: Applying the Framework to Go}\label{section:description:case-study}
Having shown how the framework captures known attacks, we next illustrate how it can guide the analysis of a programming language. Such an analysis proceeds in three steps. First, we identify candidate shared runtime state and determine whether it is mutable. Second, we determine whether attacker-controlled data can influence a Fetch from or Store to that state. If such an interaction exists, the final step is to determine whether a gadget can propagate a fetched value to a security-sensitive sink. This decomposition allows candidate attack mechanisms to be considered independently of language-specific vulnerability patterns.

Prior work has instead focused on specific form of shared runtime state.
Dahse et al.~\cite{dahse2014code} recognized magic methods are an obscure-yet-dangerous gateway to a runtime's vtable.
Lekies et al.~\cite{lekies2017code} recognized developers use the DOM to store state, demonstrating how standard DOM APIs may inadvertently read user-generated content, enabling exploitation.
Arteau~\cite{arteau2018prototype} identified prototype objects in JavaScript as shared state and cataloged how it can be accessed and changed.
Khodayari and Pellegrino~\cite{khodayari2023dom} systematized named properties in browsers based on empirical experiments.
And Liu et al.~\cite{liu2026first} systematized the language primitives enabling class pollution in Python.

We apply this process to Go. The attack classes considered in \Cref{section:description:known-attacks} rely on mechanisms such as inheritance, prototypes, object traversal, or the DOM, which do not directly transfer to Go. We therefore start from the framework rather than from an existing attack pattern and consider package-level variables as a concrete form of mutable shared state. A Go package may contain types, functions, constants, and variables that are accessible to its users. While the former are fixed at runtime, package variables are mutable, namely changing the value of a package variable in one part of the code is observable through subsequent uses of the same variable in other parts.

To assess whether this state could support a code reuse weakness, we inspected exported package variables in the Go 1.24.0 standard library.
We identified two security-relevant examples, \texttt{Reader} in \texttt{crypto/rand} and \texttt{SocketDisableIPv6} in \texttt{syscall}. Attacker control of these variables can lead to cryptographic failures and denial of service, respectively. The remaining question is whether attacker-controlled data can dynamically select such package variables for a Fetch or Store. In standard Go syntax, package members are referred to using statically defined identifiers and the reflection API does not operate on package references. In terms of the framework, this suggests Go lacks Fetch and Store primitives that can be leveraged.

This case study illustrates an important use of the framework: mutable shared state alone is insufficient for a code reuse weakness. The language must additionally provide a mechanism through which attacker-controlled data can influence how that state is accessed or modified. Our analysis therefore rules out package-variable selection as a candidate mechanism under the language features considered here.

\section{Deriving Attack Categories}\label{section:categorization}
\Cref{section:description} showed that known attacks instantiate common paths through the framework. We now use the structure of the framework to derive a categorization of weaknesses and gadgets. For weaknesses, we consider two dimensions: (1) the sequence of Fetch and Store operations leading to the final Fetch, and (2) which parameters of these operations depend on attacker input. For gadgets, we characterize how the value exposed by the final Fetch propagates to a security-sensitive sink. This analysis yields attack structures beyond those exhibited by the known attacks in \Cref{section:description}.

\subsection{Weaknesses}\label{section:categorization:weakness}
To start, we focus on categorizing code reuse attacks based on the weakness. We do this along two axis, the nodes visited when tracing paths through the weakness subgraph of \Cref{figure:framework} as well as the instantiations of the edges taken. The former allows us to describe different attacks strategies at a high level while the latter enables finer-grained classification of strategies.

\begin{figure}[]
\centering
\begin{lstlisting}[caption={A code reuse attack with a recycling weakness (JavaScript). \texttt{obj1}, \texttt{obj2}, and \texttt{obj3} are plain JavaScript objects.},label={listing:recycle}]
var i1 = "toString", i2 = "__proto__", i3 = "p"
var y = obj1[i1]
obj2[i2][i3] = y
var x = obj3.p
if (x && x()) sink()
\end{lstlisting}

\begin{lstlisting}[caption={A minimal prototype-based stateless code reuse attack (JavaScript). \texttt{obj} is a plain JavaScript object.},label={listing:stateless-prototype}]
var i1 = "toString"
var x = obj[i1]
if (x && x()) sink()
\end{lstlisting}

\begin{lstlisting}[caption={A mutating code reuse attack with a fixed value (JavaScript). \texttt{obj1} and \texttt{obj2} are plain JavaScript objects.},label={listing:mutate-trusted}]
var i1 = "__proto__", i2 = "p"
obj1[i1][i2] = () => "..."
var x = obj2.p()
eval(x)
\end{lstlisting}
\end{figure}

\subsubsection{Strategies}\label{section:categorization:strategies}
Let us start by tracing paths through the weakness subgraph of \Cref{figure:framework}. In doing so we observe three distinct strategies (we discuss extended cycles in Appendix~\ref{section:categorization-extended}):
\begin{equation*}
  \begin{aligned}
    \textit{Stateless} \; : & \quad \texttt{Data} \to \texttt{Fetch} \\
    \textit{Mutating}  \; : & \quad \texttt{Data} \to \texttt{Store} \to \texttt{Fetch} \to \ldots \\
    \textit{Recycling} \; : & \quad \texttt{Data} \to \texttt{Fetch} \to \texttt{Store} \to \texttt{Fetch} \to \ldots \\
  \end{aligned}
\end{equation*}
These paths end at the final Fetch of the weakness and the gadget begins with the value exposed by that Fetch. In \Cref{section:description:conclusion} we already noted the distinction between stateless (e.g., object injection) and stateful (e.g., prototype pollution) weaknesses.
The latter occurs when attacker influence persists through a Store to shared runtime state before the final Fetch. We further refine the stateful category into \emph{mutating} and \emph{recycling} weaknesses. In a mutating weakness, attacker input directly influences the Store. In a recycling weakness, a value first obtained from shared runtime state through a Fetch is subsequently stored into shared runtime state before being fetched again. The stateful attack examples seen thus far are all mutating.

To illustrate recycling, consider the example in \Cref{listing:recycle}. The attacker controls a property accessed on an object, picking \texttt{toString} from the root object prototype. The program assigns this to a nested property of the attacker's choosing, using it to pollute the property \texttt{p}. The program then accesses \texttt{p} on an unrelated object for a control-flow decision, reading the polluted value (i.e., the \texttt{toString} function) and causing a branch performing a sensitive operation to be taken. The first Fetch selects the pre-existing \texttt{toString} function from the prototype, the Store places this fetched value at property \texttt{p}, and the final Fetch obtains it through \texttt{obj3.p}. The attack therefore reuses a value already present in shared runtime state rather than supplying a new payload.

This demonstrates how prototype objects, considered only for mutating code reuse heretofore, can be used in a recycling attack. Similarly, prototype objects can be used in a stateless attack (this has been observed in the wild, see CVE-2025-55182), as shown \Cref{listing:stateless-prototype}. Here, pre-existing state from the root object prototype is accessed based on attacker input, resulting in a branch performing a sensitive operation to be taken.
These examples are not specific to prototypes. The same framework paths can arise whenever a runtime exposes compatible Fetch and Store mechanisms over shared state.

\subsubsection{Attacker Control}\label{section:categorization:control}
\begin{figure}[]
  \centering
  \includegraphics[width=.4\columnwidth]{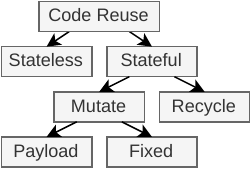}
  \caption{
    Categorization of code reuse attacks by Weakness.
  }\label{figure:categorization:weakness}
\end{figure}

We next refine weakness strategies according to what attacker-controlled data influences. We focus on mutating weaknesses, for which the relevant operations are a Store \texttt{store($l_s$, $v$)} followed by a final Fetch (\texttt{fetch($l_f$)}. This exposes three dimensions of attacker control: the location $l_s$ to which a value is stored, the value $v$ being stored, and the location $l_f$ from which a value is subsequently fetched. Different combinations of these dimensions constrain which parts of shared runtime state an attacker can manipulate and later expose to a gadget.
Appendix~\ref{section:categorization-extended} gives the complete combinations for stateless and recycling cases and
\Cref{figure:categorization:weakness} gives the full hierarchical categorization.

For the main categorization, we distinguish mutating weaknesses according to whether the attacker controls the stored value $v$. We call a weakness a \emph{payload mutation} when $v$ is attacker-controlled, directly or indirectly, and so is its store or fetch location. In contrast, in a \emph{fixed-value mutation}, $v$ is supplied by the program rather than by the attacker. For the latter, exploitation only arises because attacker-controlled data influences where that program-provided value is stored to or fetched from.

At first sight, removing attacker control over $v$ appears to make exploitation substantially less likely. However, as observed with stateless and recycling attacks, a fixed-value mutation can remain exploitable when the program itself supplies a security-relevant value. \Cref{listing:mutate-trusted} illustrates this case: the stored value is a pre-existing function value chosen by the program, while attacker-controlled property names determine where it is stored. A later Fetch exposes this function to the gadget, which invokes it and propagates its result to the sink.

Thus, control over $v$ distinguishes whether an attacker injects a payload into shared runtime state or instead repositions a program-provided value. Control over $l_s$ and $l_f$ is orthogonal since it determines which Store and Fetch locations can participate in the weakness, rather than increasing the attacker options.

\subsection{Gadgets}\label{section:categorization:gadget}
We next categorize code reuse attacks based on the gadget, i.e., how the value exposed by the final Fetch of the weakness is propagated to a security-sensitive sink. To capture gadgets that access shared runtime state again, we additionally consider the edges $\texttt{Data flow} \to^{e} \texttt{Fetch}$, $\texttt{Control flow} \to^{e} \texttt{Fetch}$, and $\texttt{Call flow} \to^{e} \texttt{Fetch}$, shown in the extended framework in \Cref{figure:framework-extended}. Tracing paths through the gadget subgraph yields three categories. Based on the terminology of~\cite{liu2024undefined} we call these categories:
\begin{equation*}
  \begin{aligned}
    \textit{Simple}           \; : & \quad \texttt{Data flow} \to \ldots \to     \texttt{Sink}                 \\
    \textit{Vertical chain}   \; : & \quad \texttt{Data flow} \to \ldots \to     \texttt{Data flow} \to \ldots \\
    \textit{Horizontal chain} \; : & \quad \texttt{Data flow} \to \ldots \to^{e} \texttt{Fetch}     \to \ldots \\
  \end{aligned}
\end{equation*}
A \textit{simple gadget} reaches the sink without revisiting Data flow or performing another Fetch. A \textit{vertical chain} revisits Data flow, for example when a fetched value influences control or call flow that subsequently produces another value. A \textit{horizontal chain} performs an additional Fetch from shared runtime state. Horizontal chains may therefore contain vertical chains and both always end in a simple gadget. When multiple categories apply, we classify a gadget as horizontal, then vertical, then simple. \Cref{listing:vertical-gadget-chain} and~\ref{listing:horizontal-gadget-chain} illustrate vertical and horizontal chains, respectively.

To obtain a finite set of canonical patterns, we enumerate non-repeating gadget paths, in which each framework edge is traversed at most once before reaching the sink. This yields 4 simple, 9 vertical, and 4 base horizontal patterns, listed in Appendix~\ref{section:categorization-extended}. Vertical chains capture structured data propagation within a gadget, whereas horizontal chains capture composition across multiple shared-state accesses.

Horizontal chains further differ in how their Fetches are related. In Listing~\ref{listing:horizontal-gadget-chain}, the first Fetch controls whether the second Fetch occurs, forming a control-dependent horizontal chain. In Listing \Cref{listing:horizontal-data-dependency}, by contrast, the value obtained by the first Fetch determines the location of the second Fetch (\texttt{obj2[x]}), forming a data-dependent horizontal chain. This illustrates how one gadget can influence not only the execution of another, but also which shared-state value it accesses.

\begin{figure}[]
\centering
\begin{lstlisting}[caption={An example vertical prototype pollution gadget chain (JavaScript). \texttt{obj} is a plain JavaScript object.},label={listing:vertical-gadget-chain}]
Object.prototype.p = { q: PAYLOAD }
var i1 = "p"
var x = obj[i1]
if (x) { var y = x.q; eval(y) }
\end{lstlisting}

\begin{lstlisting}[caption={An example horizontal prototype pollution gadget chain (JavaScript). \texttt{obj1} and \texttt{obj2} are plain JavaScript objects.},label={listing:horizontal-gadget-chain}]
Object.prototype.p = true; Object.prototype.q=PAYLOAD
var i1 = "p", i2 = "q"
var x = obj1[i1]
if (x) { var y = obj2[i2]; eval(y) }
\end{lstlisting}

\begin{lstlisting}[caption={A data-dependent horizontal gadget chain (JavaScript). \texttt{obj1} and \texttt{obj2} are plain JavaScript objects.},label={listing:horizontal-data-dependency}]
Object.prototype.p = "q"; Object.prototype.q=PAYLOAD
var i1 = "p"
var x = obj1[i1]
var y = obj2[x]
eval(y)
\end{lstlisting}
\end{figure}

In summary, the framework yields three canonical weaknesses (stateless, mutating, and recycling) with mutating weaknesses further distinguished as payload or fixed-value mutations. For gadgets, we identify simple, vertical, and horizontal structures, yielding 17 canonical patterns. These categories expose attack structures beyond the known attacks in \Cref{section:description} and provide a basis for reasoning about which parts of an attack can be restricted or eliminated. \Cref{section:defenses} uses this perspective to study defenses against code reuse attacks.

\section{Defenses}\label{section:defenses}
The framework suggests two complementary ways to reason about defenses. A defense can prevent or restrict the weakness, or it can prevent the value exposed by the weakness from reaching a security-sensitive sink through a gadget. We first revisit defenses proposed for the five attack classes in \Cref{section:description:known-attacks} and identify where they intervene in the framework. We then derive three distinct defense strategies, mediate, minimize, and constrain, and use them to identify additional defense directions.

\subsection{Mapping Known Defenses}\label{section:defenses:known-defenses}

\subsubsection{Object Injection}\label{section:defenses:object-injection}
Object injection relies on attacker influence over runtime type or method resolution, corresponding to the stateless weakness $\texttt{Data}\rightarrow\texttt{Fetch}$ in \Cref{figure:framework}. Existing defenses primarily mediate or reduce the set of types that may be selected. Type restrictions can narrow this set, while look-ahead object parsing enforces an allowlist before potentially dangerous deserialization callbacks execute~\cite{seacord2017combating}. Type restrictions alone do not prevent callbacks, e.g., magic methods~\cite{dahse2014code}, that execute before the restriction is applied. Alternatively, \changed{debloating reduces the reusable behavior available to potential gadgets}{debloating may reduce reusable behavior available to potential gadgets}~\cite{azad2019less}.

\subsubsection{Prototype Pollution}\label{section:defenses:prototype-pollution}
Prototype pollution follows the mutating weakness $\texttt{Data}\rightarrow\texttt{Store}\rightarrow\texttt{Fetch}$ in \Cref{figure:framework}. Property filtering mediates attacker-controlled navigation to prototype state, while freezing or restricting extensibility prevents its mutation~\cite{cornelissen2024universal}. The proposed Symbol.proto mechanism instead restricts access to the prototype hierarchy through stratification~\cite{symbolprotoproposal}. Defenses such as \texttt{Object.hasOwn}, \texttt{Map}s, and prototype-free objects prevent subsequent property lookups from exposing polluted prototype state to gadgets~\cite{cornelissen2024universal}.

\subsubsection{Class Pollution}\label{section:defenses:class-pollution}
Class pollution similarly follows a mutating weakness: attacker-controlled metaprogramming operations navigate to shared class or module state, modify it, and a later Fetch exposes the modification. Filtering values used in dynamic accesses such as \texttt{getattr} can restrict this navigation~\cite{liu2026first}. Among the defenses considered here, we did not identify a general gadget defense that acts after the polluted value has been fetched.

\subsubsection{DOM Clobbering}\label{section:defenses:dom-clobbering}
DOM clobbering relies on a mutating weakness in which attacker-controlled DOM content affects browser-managed named-property state that is subsequently fetched through a property lookup. Weakness-side defenses include removing or namespacing \texttt{name} and \texttt{id} attributes, restricting named-property behavior, and protecting individual properties from shadowing~\cite{janc2020oh,khodayari2023dom}. After the Fetch, type validation and Content Security Policy can prevent the resulting value from reaching a sensitive operation~\cite{khodayari2023dom}.

\subsubsection{DOM Script Gadgets}\label{section:defenses:dom-script-gadgets}
DOM script gadgets also rely on a mutating weakness, but storing untrusted content in the DOM is often intentional. Defenses therefore commonly act after the Fetch by restricting how DOM-derived values are propagated. Sanitization, Content Security Policy, Trusted Types, and validation of values read from the DOM can prevent particular flows to sensitive operations~\cite{drescher2025dom}.

\begin{table}
  \centering
  \caption{
    Existing defense approaches categorized into three high-level strategies.
    Concrete examples in parenthesis.
    A complete mapping of concrete defenses is provided in \Cref{table:defenses-3}.
  }\label{table:defenses-1}
  \begin{tabular}{ c | l }
    \textbf{Mediate}   & Stratification (Symbol.proto) Allow/Blocklist    \\
                       & (Look-ahead object parsing), Namespace (Prefix)  \\
                       & Validation (Type check), Abstraction (Map)       \\
      \midrule
    \textbf{Minimize}  & Remove (Debloating), Avoid (Disable named        \\
                       & properties)                                      \\
      \midrule
    \textbf{Constrain} & Immutability (Object.freeze), Extensibility      \\
                       & (Object.seal)                                    \\
  \end{tabular}
\end{table}

\subsection{Defense Strategies}\label{section:defenses:insights}
The defenses considered above instantiate three recurring strategies: mediate, minimize, and constrain. These strategies are complementary to the weakness–gadget distinction. They characterize how a defense restricts an attack, while the framework identifies where that restriction takes effect. \Cref{table:defenses-1} groups the concrete defense approaches according to these strategies and \Cref{table:defenses-3} provides the detailed mapping to the framework.

\emph{Mediate} controls whether and how shared runtime state can be accessed, modified, or propagated. This includes stratification~\cite{bracha2004mirrors,symbolprotoproposal}, allow- and blocklisting, namespacing, validation, and the use of alternative abstractions. Depending on where mediation is applied, it can restrict attacker-controlled Fetch or Store operations in the weakness, prevent a stored value from being exposed by the final Fetch, or prevent a fetched value from propagating through a gadget.

\emph{Minimize} reduces the attack surface available for code reuse. This may involve removing unused state or behavior, as with debloating, or avoiding mechanisms that expose shared runtime state, such as named properties. Minimization can therefore apply to both weaknesses and gadgets: reducing available shared state limits attacker-controlled interactions, while reducing reusable behavior limits the operations available to gadgets.

\emph{Constrain} restricts how shared runtime state may change after it has been created. Immutability and restricted extensibility are instances of this strategy. Constraining primarily targets stateful weaknesses, since both mutating and recycling weaknesses require a Store before the final Fetch, whereas stateless weaknesses do not modify shared state.

These strategies provide a structure to reason about our Go case study (\Cref{section:description:case-study}). In particular, Go mediates access to shared state through stratification and an intentionally-scoped reflection API, and has minimal amount of shared runtime state--explaining why we did not identify a code reuse attack class for Go. Yet, constrainment can be applied by using package constants over variables.

\begin{table}
  \centering
  \caption{
    Defense approaches for the five attack classes.
    \emptycirc~means no known use, \fullcirc~means a known use, and \halfcirc~means framework-derived candidates proposed in this work.
  }\label{table:defenses-2}
  \begin{tabular}{ r c c c c c }
                             & \textbf{OI} & \textbf{PP} & \textbf{CP} & \textbf{DC} & \textbf{\changed{DS}{DSG}} \\
      \midrule
    \textbf{Stratification}  & \emptycirc  & \fullcirc   & \halfcirc   & \emptycirc  & \emptycirc  \\
    \textbf{Allow/Blocklist} & \fullcirc   & \fullcirc   & \fullcirc   & \fullcirc   & \emptycirc  \\
    \textbf{Namespace}       & \emptycirc  & \emptycirc  & \emptycirc  & \fullcirc   & \emptycirc  \\
    \textbf{Validation}      & \emptycirc  & \fullcirc   & \emptycirc  & \fullcirc   & \fullcirc   \\
    \textbf{Abstraction}     & \emptycirc  & \fullcirc   & \emptycirc  & \halfcirc   & \emptycirc  \\
      \midrule
    \textbf{Remove}          & \fullcirc   & \emptycirc  & \emptycirc  & \emptycirc  & \emptycirc  \\
    \textbf{Avoid}           & \halfcirc   & \halfcirc   & \emptycirc  & \fullcirc   & \halfcirc   \\
      \midrule
    \textbf{Immutability}    & \emptycirc  & \fullcirc   & \emptycirc  & \fullcirc   & \emptycirc  \\
    \textbf{Extensibility}   & \emptycirc  & \fullcirc   & \emptycirc  & \emptycirc  & \emptycirc  \\
  \end{tabular}
\end{table}

\subsection{Framework-Derived Defense Directions}\label{section:defenses:new-defenses}
The defense strategies also expose combinations of attack classes and defense approaches that, among the defenses considered here, have not previously been applied. Table 2 highlights these gaps and motivates several candidate defense directions. These are not evaluated defenses but they illustrate how strategies used against one attack class can suggest defenses for another.

For object injection, minimization suggests reducing the security-sensitive behavior available through attacker-influenced dynamic dispatch. In particular, sensitive operations could be separated from classes that may be selected through untrusted deserialization and instead be exposed through statically referenced operations where dynamic dispatch is unnecessary.

For prototype pollution attacks, Listings \ref{listing:recycle}, \ref{listing:stateless-prototype}, and \ref{listing:mutate-trusted} show that function values exposed through shared prototype state can participate in stateless, mutating, and recycling weaknesses. Minimization therefore suggests reducing such callable behavior, for example by moving security-sensitive operations out of the prototype chain when dynamic instance dispatch is unnecessary.
For class pollution, \Cref{section:description:class-pollution} shows that attacker-controlled dynamic attribute accesses can traverse from application objects through meta-level attributes into shared class or module state. Mediation suggests applying a stratification approach a la Symbol.proto~\cite{symbolprotoproposal}, separating meta-level attribute from ordinary dynamic attribute access.

For DOM clobbering, mediation suggests encapsulating accesses to clobberable DOM namespaces behind abstractions that perform explicit lookup and validation rather than relying directly on named-property resolution. This would focus existing defensive checks in a reusable interface instead of requiring individual gadgets to apply them correctly.
For DOM script gadgets, minimization suggests separating application state from attacker-influenced DOM state. This removes the need to fetch data from the DOM, avoiding the creation of DOM script gadgets altogether.

In summary, the framework and defense strategies separate where a defense intervenes from how it restricts an attack. Weakness and gadget defenses identify the point of intervention, while mediation, minimization, and constraint characterize the strategy applied there. The candidate directions above illustrate how this separation can be used to transfer defense principles across attack classes.

\section{Modeling Language}\label{section:model}
We now give a formal account of the framework of \Cref{section:description:framework}. To this end, we define a simple imperative language $\modellanguagename$, with explicit shared runtime state and observable Fetch and Store operations. Our formalization focuses on attacker influence over which shared state locations are accessed or modified, deliberately abstracting from changes that affect only values stored at fixed locations. We use the model to define code reuse weaknesses, characterize gadgets and end-to-end vulnerabilities, and encode the five attack classes from \Cref{section:description:known-attacks}.

\subsection{$\modellanguagename$}\label{section:model:language}
In the grammar of $\modellanguagename$, presented in \Cref{figure:model:grammar}, the metavariable $e$ ranges over expressions, $n$ over integers, $x$ over variables, and $c$ over commands. We use the metavariable $l$ for locations (split between heap, $l_o$, and function, $l_f$). Values $v$ are locations, numbers, or $\nil$. The semantics enforces memory safety by distinguishing locations from integers, allowing heap accesses only through valid heap locations and integer indices.

An expression is either a number, a variable, nil ($\nil$), or a binary operation ($\binop$) over expressions.
Programs consist of commands, including the standard skip, sequence ($c_1; c_2$), assignment ($x := e$), if, and while. We add $x \gets \kwalloc$ to allocate a new object and $x := x_o[e_i]$ and $x[e_i] := e$ for reading from and writing to objects. We use $\kwfn x \;\{ c \}$ for user-defined functions and $\kwcall x$ to call them.

\begin{figure}[]
  \begin{equation*}
    \begin{aligned}
      e ::=&\; \nil \mid n \mid x \mid e_1 \binop e_2                                     & \text{Expressions} \\
      c ::=&\; \kwskip \mid c_1; c_2 \mid x := e \mid x[e_i] := e \mid x := x_o[e_i] \mid & \text{Commands}    \\
           &\; x \gets \kwalloc \mid \kwfn x \kwblock{c} \mid \kwcall x \mid              &                    \\
           &\; \kwif e \kwblock{c_1} \kwelse \kwblock{c_2} \mid \kwwhile e \kwblock{c}    &                    \\
    \end{aligned}
  \end{equation*}
  \caption{
    Grammar for $\modellanguagename$.
  }\label{figure:model:grammar}
\end{figure}

\subsubsection{Operational Semantics}\label{section:model:semantics}
We give $\modellanguagename$ a small-step operational semantics, see \Cref{figure:semantics}. Configurations consist of a command $c$, store $\Store : x \to v$, heap $\Heap : l_o \to n \to v$, function namespace $\Functions : l_f \to c$, and runtime context $\Context : x \to l_o$. The stores associates variables with values (\textsc{E-Assign}). The heap associates locations with objects (\textsc{E-Alloc}). The function namespace associates function locations with function bodies (\textsc{E-Func}). The runtime context stores heap locations associated with variables (\textsc{E-Alloc-C}). The latter models the runtime state as discussed in \Cref{section:description}, which we will utilize in \Cref{section:model:security-condition}.

Expression evaluation, $[\![e]\!]_{\State}$, is defined as
\begin{equation*}
  \begin{aligned}
    [\![v]\!]_{\State}              &= v                                                \\
    [\![x]\!]_{\State}              &= \State(x)                                        \\
    [\![e_1 \binop e_2]\!]_{\State} &= [\![e_1]\!]_{\State} \binop [\![e_2]\!]_{\State} \\
  \end{aligned}
\end{equation*}
where
\begin{equation*}
  \State(x) \isdefinedas
    \begin{cases}
      \; \Store(x)     & \mathit{if} \; x \in \Store     \\
      \; \Context(x)   & \mathit{if} \; x \not\in \Store \\
    \end{cases}
\end{equation*}
and\footnote{The semantics of $\land$ are analogous to the \kwif command.}
\begin{equation*}
  \binop ::=
    \begin{cases}
      \; + \mid - \mid \land & \mathit{if} \; [\![e_1]\!]_{\State} = n_1 \land [\![e_2]\!]_{\State} = n_2 \\
      \; = \;\mid\; \neq     & \mathit{if} \; [\![e_1]\!]_{\State} = v_1 \land [\![e_2]\!]_{\State} = v_2 \\
    \end{cases}
\end{equation*}

\begin{figure*}[t]
  \centering
  \begin{mathpar}
      \frac{
        l_o = \State(x)
          \quad
        l_o \in \locations_{\Heap}
          \quad
        n = [\![e_i]\!]_{\State}
          \quad
        n \in \numbers
          \quad
        v = [\![e]\!]_{\State}
      }{
        \langle x[e_i] := e, \Store, \Heap, \Functions, \Context \rangle
          \stackrel{\observation^w_{l_o,n}}{\step}
        \langle \epsilon, \Store, \Heap(l_o)[n \mapsto v], \Functions, \Context \rangle
      } \; \text{(\textsc{E-Store})}
      \quad
      \frac{
        l_o = \State(x_o)
          \quad
        l_o \in \locations_{\Heap}
          \quad
        n = [\![e_i]\!]_{\State}
          \quad
        n \in \numbers
          \quad
        v = \Heap(l_o, n)
      }{
        \langle x := x_o[e_i], \Store, \Heap, \Functions, \Context \rangle
          \stackrel{\observation^r_{l_o,n}}{\step}
        \langle \epsilon, \Store[x \mapsto v], \Heap, \Functions, \Context \rangle
      }\; \text{(\textsc{E-Fetch})}

      \frac{
        l_o \not\in \Heap
      }{
        \langle x \gets \kwalloc, \Store, \Heap, \Functions, \Context \rangle
          \step
        \langle \epsilon, \Store[x \mapsto l_o], \Heap[l_o \mapsto \emptyset], \Functions, \Context \rangle
      } \; \text{(\textsc{E-Alloc})}
      \quad
      \frac{
        l_o \not\in \Heap
      }{
        \langle x \gets \kwalloc_{\Context}, \Store, \Heap, \Functions, \Context \rangle
          \step
        \langle \epsilon, \Store, \Heap[l_o \mapsto \emptyset], \Functions, \Context[x \mapsto l_o] \rangle
      } \; \text{(\textsc{E-Alloc-C})}
  \end{mathpar}
  \caption{
    Operational semantics \changed{of}{for the} unique parts of $\modellanguagename$,
    remaining commands are straightforward (see \Cref{figure:semantics-rest}).
  }\label{figure:semantics}
\end{figure*}

The evaluation relation $\stackrel{\observation}{\step}$ is defined by \Cref{figure:semantics}. The annotation $\observation$ is used to capture \emph{observations} of heap accesses and is omitted when it is not applicable. Observations are only produced by \textsc{E-Store} and \textsc{E-Fetch}, recording if it is a read ($\observation^r$) or write ($\observation^w$) and the location ($\observation_{l}$) and index ($\observation_{\cdot,n}$).%

Observations intentionally do not record the value read or written. We are interested in whether attacker input influences part of shared runtime state that is accessed or modified. Changes that affect only the value written to a fixed location correspond to ordinary data flow and are outside our weakness abstraction.

Note that the semantics can get stuck in various situations, such as when accessing the heap using a non-location value or adding a number to a location.

\subsubsection{Input and Output}
We model inputs and outputs through reserved variable names.
In particular, inputs are modeled as the variables $\Input_i$ ($i \in \mathbb{N}$) with a numeric value in $\Store$
and output is modeled by the value of the $\Output$ variable in $\Store$ at the end of an execution.

\subsection{Security Conditions}\label{section:model:security-condition}
We will now define a security condition that captures when a program contains a code reuse attack \emph{vulnerability}.
To this end, we define a trace $\trace$ as multiset (notation: $\multiset{\cdots}$) of observations collected during an execution (a sequence of transitions in the transitive closure of $\stackrel{\observation}{\step}$). We deliberately abstract from their order because changing only the order in which the same shared-state locations are accessed is not a weakness under our definition (e.g., \Cref{listing:model:observation-order}). Together with the omission of accessed values from observations, this makes the security condition sensitive only to attacker influence over which shared-state locations are accessed or modified.

Next, we define the shared runtime state as all heap locations reachable from the runtime context roots in $\Context$. Intuitively, a location $l$ is reachable if there exists a sequence of pointers originating in $\Context$ and ending in $l$. Formally,
\begin{equation*}
  \begin{aligned}
    \sharedmem &\isdefinedas \{ l \mid l \in \Heap \land l \in R(C) \}                             \\
    R(o)       &\isdefinedas L(o) \cup \bigcup_{l \in L(o)} \mathit{R}(\Heap(l))                   \\
    L(o)       &\isdefinedas \{ l \mid \exists \alpha \in o. \; l = o(\alpha) \land l \in \Heap \} \\
  \end{aligned}
\end{equation*}
For a terminating execution, we determine the shared runtime state with respect to its final heap and runtime context. This includes initialization of objects that subsequently become reachable from runtime state.

Next, we define when two stores are similar. Intuitively this is when the same set of attacker-controlled input variables is defined, regardless of value, and all other defined variables have identical values. Formally,
\begin{equation*}
  \begin{aligned}
    \Store_1 \simeq \Store_2 \Leftrightarrow \forall x \in \variables. \;
      &(x \in \{ \Input_i \mid i \in \mathbb{N} \} \land \Store_1(x) \in \mathbb{Z} \land \Store_2(x) \in \mathbb{Z}) \\
      &\lor (\Store_1(x) = \Store_2(x)) \lor (x \not\in \Store_1 \land x \not\in \Store_2)                            \\
  \end{aligned}
\end{equation*}
Lastly, let $\trustedmem$ be a program-specific allowlist of shared-state locations whose use is permitted by the intended program behavior. We use it to distinguish intended uses of shared mechanisms from attacker-induced access of unintended shared state. We define the latter as the \emph{untrusted} shared runtime state $\sharedmem^u = \sharedmem \setminus \trustedmem$.
Now we can define what it means for a program to have a code reuse weakness.

Intuitively, a code reuse weakness exists when attacker-controlled input can influence which untrusted shared runtime state the program accesses or modifies. We express this as an information-flow property over two-executions. Starting from states that differ only in attacker inputs, we compare the Fetch and Store observations on shared runtime state. If these differ, then attacker input has influenced the selection or navigation of shared runtime state.
Formally,
\begin{definition}\label{definition:code-reuse-weakness}
  Given $\trustedmem$, a program $\Program$ has a code reuse \emph{weakness} if there exist a pair of stores $\Store_1 \simeq \Store_2$, heap $\Heap$, function namespace $\Functions$, and runtime context $\Context$ s.t.
  \begin{equation*}
    \begin{aligned}
      \langle \Program, \Store_1, \Heap, \Functions, \Context \rangle &\stackrel{\tau_1}{\step _{*}} \langle \epsilon, \Store_1', \Heap_1', \Functions_1', \Context_1' \rangle \\
      \langle \Program, \Store_2, \Heap, \Functions, \Context \rangle &\stackrel{\tau_2}{\step _{*}} \langle \epsilon, \Store_2', \Heap_2', \Functions_2', \Context_2' \rangle \\
    \end{aligned}
  \end{equation*}
  and\footnote{\changed{We assume the existence of a bijection $\beta$ to relate heap locations between the two executions.}{We assume the existence of a bijection relating heap locations between two executions.}}
  \begin{equation*}
    \trace_1 \cap \sharedmem^u_1 \neq \trace_2 \cap \sharedmem^u_2,
  \end{equation*}
  where
  \begin{equation*}
    \trace \cap \sharedmem^u \isdefinedas \multiset{\; \observation \mid \observation_{l} \in \trace \land l \in \sharedmem^u \;}.
  \end{equation*}
\end{definition}
Because observations ignore accessed values, this definition does not classify an execution pair as a weakness when attacker input changes only the value stored at an otherwise fixed shared-state location.
Next, we define a security condition for gadgets.

Intuitively, following \Cref{section:description:framework}, if a value from shared runtime state is used by the program for a security sensitive operation, then it has a code reuse gadget. We capture this as an information flow property as well. Namely, when there is any execution of the program for which the value fetched from the shared runtime state causes a change in its output, it must mean the program performs a sensitive operation based on it.
Formally,
\begin{definition}\label{definition:code-reuse-gadget}
  Given $\trustedmem$, a program $\Program$ has a potential code reuse \emph{gadget} if there exist a store $\Store$, pair of heaps $\Heap, \Heap'$, pair of function namespaces $\Functions, \Functions'$, runtime context $\Context$, observation $\observation^r_{l,n}$ (s.t. $l \in \sharedmem^u$), and values $v_1, v_2 \in \mathbb{Z} \cup \Heap' \cup \Functions'$ s.t.
  \begin{equation*}
    \begin{aligned}
    \langle \Program, \Store, \Heap, \Functions, \Context \rangle \stackrel{\tau}{\step _{*}} \langle \Program', \Store', \Heap', \Functions', \Context' \rangle                    &\stackrel{\observation^r_{l,n}}{\step} \langle \Program'', \Store'', \Heap'', \Functions'', \Context'' \rangle       \\
                                                                                              \langle \Program', \Store', \Heap'(l)[n \mapsto v_1], \Functions', \Context' \rangle  &\stackrel{\tau_1}{\step _{*}}          \langle \epsilon, \Store_1'', \Heap_1'', \Functions_1'', \Context_1'' \rangle \\
                                                                                              \langle \Program', \Store', \Heap'(l)[n \mapsto v_2], \Functions', \Context' \rangle  &\stackrel{\tau_2}{\step _{*}}          \langle \epsilon, \Store_2'', \Heap_2'', \Functions_2'', \Context_2'' \rangle \\
    \end{aligned}
  \end{equation*}
  and
  \begin{equation*}
    \Store_1''(\Output) \neq \Store_2''(\Output).
  \end{equation*}
\end{definition}
This definition is an overapproximation because it asks whether changing the result of a Fetch can affect the sink, without requiring both substituted values to arise in feasible executions. \Cref{listing:model:definition-2-overapproximation} illustrates the resulting false positive.
Next we combine the two definitions to define a code reuse vulnerability.

Intuitively, a program has a potential code reuse vulnerability if it has both a weakness and a gadget. While this is necessary, it may be insufficient. This is because the weakness and gadget may not be related. We approximate the relation by requiring that (1) the initial configuration of the gadget corresponds to an initial configuration for which the program has a weakness and (2) there is a weakness observation associated with a memory location for which there is a gadget.
Formally,
\begin{definition}\label{definition:code-reuse-vulnerability}
  Given $\trustedmem$, a program $\Program$ has a potential code reuse \emph{vulnerability} if there exist $\Store_1 \simeq \Store_2$, $\Heap$, $\Functions$, and $\Context$ s.t.
  \begin{equation}\label{equation:1}
    \begin{aligned}
      \langle \Program, \Store_1, \Heap, \Functions, \Context \rangle &\stackrel{\tau_1}{\step _{*}} \langle \epsilon, \Store_1', \Heap_1', \Functions_1', \Context_1' \rangle \\
      \langle \Program, \Store_2, \Heap, \Functions, \Context \rangle &\stackrel{\tau_2}{\step _{*}} \langle \epsilon, \Store_2', \Heap_2', \Functions_2', \Context_2' \rangle \\
      \trace_1 \cap \sharedmem^u_1 &\neq \trace_2 \cap \sharedmem^u_2, \\
    \end{aligned}
  \end{equation}
  and there exists a $\Heap'$, $\Functions'$, $\observation^r_{l,n}$, and $v_1, v_2 \in \mathbb{Z} \cup \Heap' \cup \Functions'$ s.t.
  \begin{equation}\label{equation:2}
    \begin{aligned}
    \langle \Program, \Store_i, \Heap, \Functions, \Context \rangle \stackrel{\tau}{\step _{*}} \langle \Program', \Store', \Heap', \Functions', \Context' \rangle                    &\stackrel{\observation^r_{l,n}}{\step} \langle \Program'', \Store'', \Heap'', \Functions'', \Context'' \rangle       \\
                                                                                                \langle \Program', \Store', \Heap'(l)[n \mapsto v_1], \Functions', \Context' \rangle  &\stackrel{\tau_1'}{\step _{*}}         \langle \epsilon, \Store_1'', \Heap_1'', \Functions_1'', \Context_1'' \rangle \\
                                                                                                \langle \Program', \Store', \Heap'(l)[n \mapsto v_2], \Functions', \Context' \rangle  &\stackrel{\tau_2'}{\step _{*}}         \langle \epsilon, \Store_2'', \Heap_2'', \Functions_2'', \Context_2'' \rangle \\
    \Store_1''(\Output) &\neq \Store_2''(\Output),
    \end{aligned}
  \end{equation}
  and there exists $\observation_{l',n'} \in (\trace_1 \cap \sharedmem^u_1) \; \Delta \; (\trace_2 \cap \sharedmem^u_2)$\footnote{$\Delta$ denotes symmetric difference.} s.t.
  \begin{equation}\label{equation:3}
    l'=l \land n'=n \land \observation_{l',n'} \in \tau \cup \{ \observation^r_{l,n} \}.
  \end{equation}
\end{definition}
The condition combines a weakness witness (\ref{equation:1}) with a potential gadget (\ref{equation:2}) and requires the two to concern the same shared-state location and index (\ref{equation:3}). This establishes a structural connection between the weakness and gadget but does not ensure that their combination is feasible along one execution. Consequently, Definition 3 remains an overapproximation, as illustrated by \Cref{listing:model:definition-3-overapproximation}.

We can refine these definitions to capture stateless and stateful code reuse attacks (see \Cref{section:categorization}). To achieve this, let $\trace^w$ denote the subset of write observations $\observation^w \in \trace$. Then, a stateless weakness/vulnerability (additionally) requires $\trace^w_1 \cap \sharedmem^u_1 = \trace^w_2 \cap \sharedmem^u_2$ and a stateful weakness/vulnerability requires $\trace^w_1 \cap \sharedmem^u_1 \neq \trace^w_2 \cap \sharedmem^u_2$. The latter overapproximates, a stateful weakness does not preclude a stateless weakness and a stateless vulnerability may be misclassified as stateful (e.g., \Cref{listing:model:definition-stateless-stateful}).

\subsection{Validation against Known Attacks}\label{section:model:known-attacks}
We validate the model against the five known attack classes used to derive the framework in \Cref{section:description:known-attacks}. We encode \Cref{listing:object-injection} through~\ref{listing:dom-script-gadgets} in $\modellanguagename$ and show that the security condition captures the vulnerabilities. We present object injection and prototype pollution here and refer to \Cref{section:model-extended} for the remaining three encodings.

\begin{figure}[]
  \centering
  \begin{equation*}
    \begin{aligned}
       1 \quad & \texttt{type} := \texttt{0}; \; \texttt{exec} := \texttt{1}; \; \kwfn \texttt{f} \kwblock{\kwskip} \; \kwfn \texttt{g} \kwblock{\Output := \texttt{67}} \\
       2 \quad & \texttt{\textcolor{gray}{// Initialization}}                                                                                                            \\
       3 \quad & \texttt{Cmd} \gets \kwalloc_{\Context}; \; \texttt{Bkp} \gets \kwalloc_{\Context}; \; \texttt{OSCmd} \gets \kwalloc_{\Context};                         \\
       4 \quad & \texttt{Bkp}[\texttt{type}] := \texttt{Cmd}; \; \texttt{Bkp}[\texttt{exec}] := \texttt{f};                                                              \\
       5 \quad & \texttt{OSCmd}[\texttt{type}] := \texttt{Cmd}; \; \texttt{OSCmd}[\texttt{exec}] := \texttt{g};                                                          \\
       6 \quad & \texttt{\textcolor{gray}{// `Cmd c = (Cmd) deserialize(i1);'}}                                                                                          \\
       7 \quad & \texttt{c} \gets \kwalloc;                                                                                                                              \\
       8 \quad & \kwif \Input_1 = \texttt{0} \kwblock{\texttt{c}[\texttt{type}] := \texttt{Bkp}}                                                                         \\
       9 \quad & \!\kwelse \kwblock{\kwif \Input_1 = \texttt{1} \kwblock{\texttt{c}[\texttt{type}] := \texttt{OSCmd}} \kwelse \kwblock{\kwskip}}                         \\
      10 \quad & \texttt{\textcolor{gray}{// `Runnable x = c::exec;'}}                                                                                                   \\
      11 \quad & \texttt{x} := \nil; \; \texttt{t} := \texttt{c}; \; \texttt{p} := \texttt{exec};                                                                        \\
      12 \quad & \kwwhile (\texttt{x} = \nil) \land (\texttt{t} \neq \nil) \kwblock{\texttt{x} := \texttt{t}[\texttt{p}]; \; \texttt{t} := \texttt{t}[\texttt{type}]}    \\
      13 \quad & \texttt{\textcolor{gray}{// `x.run();'}}                                                                                                                \\
      14 \quad & \kwcall \texttt{x}                                                                                                                                      \\
    \end{aligned}
  \end{equation*}
  \captionsetup{type=lstlisting}
  \caption{
    Object injection in $\modellanguagename$.
  }\label{listing:model:object-injection}
\end{figure}

\subsubsection{Object Injection}\label{section:model:object-injection}
\Cref{listing:model:object-injection} encodes \Cref{listing:object-injection} in $\modellanguagename$.
Recall that this code allows the attacker to select a class to be instantiated and by extension pick and implementation of an interface method. This is enabled by deserialization of untrusted bytes, modeled by setting the \texttt{type} of the object \texttt{c} based on the input $\Input_1$ (line 7--9). The code finds the \texttt{exec} method of \texttt{c} (line 11--12), recursively checking its super class, before \texttt{call}ing it (line 14).

For this program we can demonstrate how the security conditions capture its exploitation. First, the input pair
\begin{equation*}
  \begin{aligned}
    &\Store_1[\Input_1 \mapsto 0] \\
    &\Store_2[\Input_1 \mapsto 1] \\
  \end{aligned}
\end{equation*}
produces the traces (identical write observations omitted for brevity)
\begin{equation*}
  \begin{aligned}
    \trace_1 &= \multiset{\cdots, \observation^r_{[\![\texttt{c}]\!],1}, \observation^r_{[\![\texttt{c}]\!],0}, \observation^r_{[\![\texttt{Bkp}]\!],1},   \observation^r_{[\![\texttt{Bkp}]\!],0}}   \\
    \trace_2 &= \multiset{\cdots, \observation^r_{[\![\texttt{c}]\!],1}, \observation^r_{[\![\texttt{c}]\!],0}, \observation^r_{[\![\texttt{OSCmd}]\!],1}, \observation^r_{[\![\texttt{OSCmd}]\!],0}} \\
  \end{aligned}
\end{equation*}
given $\trustedmem = \{ [\![\texttt{Bkp}]\!] \}$ (i.e., \texttt{Bkp} is intended to be instantiated) we get $\sharedmem^u_i = \{ [\![\texttt{Cmd}]\!], [\![\texttt{OSCmd}]\!] \}$ and find
\begin{equation*}
  \begin{aligned}
    \trace_1 \cap \sharedmem^u_1 &\neq \trace_2 \cap \sharedmem^u_2                                                                            \\
               \multiset{\cdots} &\neq \multiset{\cdots, \observation^r_{[\![\texttt{OSCmd}]\!],1}, \observation^r_{[\![\texttt{OSCmd}]\!],0}} \\
  \end{aligned}
\end{equation*}
satisfying the security condition for a code reuse weakness.
If we consider the second execution, we can execute until $\observation^r_{[\![\texttt{OSCmd}]\!],1}$ (corresponding to $\texttt{t}[\texttt{p}]$, line 12) and continue execution with $v_1=[\![\texttt{f}]\!]$ and $v_2=[\![\texttt{g}]\!]$ until completion to observe different outputs, $\nil$ and $\texttt{67}$ resp., thus satisfying the security condition for a code reuse gadget.
Since we used an initial state of the weakness to demonstrate the gadget, and the modified observation is related to an observation in the delta between the two executions prior to it, namely the observation itself, the security condition for a code reuse vulnerability is also met.

Moreover, because for the weakness we showed that $\trace^w_1 \cap \sharedmem^u_1 = \trace^w_2 \cap \sharedmem^u_2$, the definition refinement correctly classifies this as a \emph{stateless} vulnerability.

\begin{figure}[]
  \centering
  \begin{equation*}
    \begin{aligned}
       1 \quad & \texttt{\_\_proto\_\_} := \texttt{0}; \; \texttt{p} := \texttt{1};                                                                                            \\
       2 \quad & \texttt{\textcolor{gray}{// Initialization}}                                                                                                                  \\
       3 \quad & \texttt{prototype} \gets \kwalloc_{\Context};                                                                                                                 \\
       4 \quad & \texttt{obj1} \gets \kwalloc; \; \texttt{obj1}[\texttt{\_\_proto\_\_}] := \texttt{prototype};                                                                 \\
       5 \quad & \texttt{obj2} \gets \kwalloc; \; \texttt{obj2}[\texttt{\_\_proto\_\_}] := \texttt{prototype};                                                                 \\
       6 \quad & \texttt{\textcolor{gray}{// `obj1[i1][i2] = i3'}}                                                                                                             \\
       7 \quad & \texttt{t} := \texttt{obj1}[\Input_1]; \; \texttt{t}[\Input_2] := \Input_3;                                                                                   \\
       8 \quad & \texttt{\textcolor{gray}{// `var x = obj2.p || "..."'}}                                                                                                       \\
       9 \quad & \texttt{x} := \nil; \; \texttt{t} := \texttt{obj2};                                                                                                           \\
      10 \quad & \kwwhile (\texttt{x} = \nil) \land (\texttt{t} \neq \nil) \kwblock{\texttt{x} := \texttt{t}[\texttt{p}]; \; \texttt{t} := \texttt{t}[\texttt{\_\_proto\_\_}]} \\
      11 \quad & \kwif \texttt{x} = \nil \kwblock{\texttt{x} := \texttt{42}} \kwelse \kwblock{\kwskip}                                                                         \\
      12 \quad & \texttt{\textcolor{gray}{// `eval(x)'}}                                                                                                                       \\
      13 \quad & \Output := \texttt{x}                                                                                                                                         \\
    \end{aligned}
  \end{equation*}
  \captionsetup{type=lstlisting}
  \caption{
    Prototype pollution in $\modellanguagename$.
  }\label{listing:model:prototype-pollution}
\end{figure}

\subsubsection{Prototype Pollution}\label{section:model:prototype-pollution}
\Cref{listing:model:prototype-pollution} encodes \Cref{listing:prototype-pollution} in $\modellanguagename$.
Recall that this code allows the attacker to alter a prototype object through a dynamic property lookup on an object and subsequent assignment. We model object prototypes through an explicit assignment of the \texttt{\_\_proto\_\_} field (line 3--5). The code then performs a nested assignment on the first object (line 7) and reads from the second object (line 9--11), recursively checking the prototype, before outputting the obtained value (line 13).

For this program we can demonstrate how the security conditions capture its exploitation. First, the input pair
\begin{equation*}
  \begin{aligned}
    &\Store_1[\Input_1 \mapsto 0, \Input_1 \mapsto 6, \Input_1 \mapsto 7]  \\
    &\Store_2[\Input_1 \mapsto 0, \Input_1 \mapsto 1, \Input_1 \mapsto 67] \\
  \end{aligned}
\end{equation*}
produces the traces (identical observations omitted for brevity)
\begin{equation*}
  \begin{aligned}
    \trace_1 &= \multiset{\cdots, \observation^w_{[\![\texttt{prototype}]\!],6}} \\
    \trace_2 &= \multiset{\cdots, \observation^w_{[\![\texttt{prototype}]\!],1}} \\
  \end{aligned}
\end{equation*}
given $\trustedmem = \emptyset$ we have $\sharedmem^u_i = \{ [\![\texttt{prototype}]\!] \}$ and find
\begin{equation*}
  \begin{aligned}
                                        \trace_1 \cap \sharedmem^u_1 &\neq \trace_2 \cap \sharedmem^u_2                                     \\
    \multiset{\cdots, \observation^w_{[\![\texttt{prototype}]\!],6}} &\neq \multiset{\cdots, \observation^w_{[\![\texttt{prototype}]\!],1}} \\
  \end{aligned}
\end{equation*}
satisfying the security condition for a code reuse weakness.
If we consider the second execution, we can execute until $\observation^r_{[\![\texttt{prototype}]\!],1}$ (corresponding to $\texttt{t}[\texttt{p}]$, line 10) and continue execution with $v_1=\texttt{6}$ and $v_2=\texttt{7}$ until completion to observe different outputs, $\texttt{6}$ and $\texttt{7}$ resp., thus satisfying the security condition for a code reuse gadget.
Since we used an initial state of the weakness to demonstrate the gadget, and the modified observation is related to an observation in the delta between the two executions prior to it, namely $\observation^w_{[\![\texttt{prototype}]\!],1}$, the security condition for a code reuse vulnerability is also met.

Moreover, because for the weakness we showed that $\trace^w_1 \cap \sharedmem^u_1 \neq \trace^w_2 \cap \sharedmem^u_2$, the definition refinement correctly classifies this as a \emph{stateful} vulnerability.

\subsection{Discussion and Limitations}\label{section:model:conclusion}
We have given a formal characterization of code reuse weaknesses in $\modellanguagename$ and shown that it captures the known attacks in \Cref{section:description}. The definition focuses on attacker influence over which shared runtime state is accessed or modified. Observations omit accessed values and are compared as multisets: changes that affect only a stored value or only the order of accesses are intentionally ignored. The model can therefore distinguish stateless and stateful weaknesses, but does not capture the finer payload/fixed-value or mutating/recycling distinctions of \Cref{section:categorization}.

\Cref{definition:code-reuse-gadget} and \ref{definition:code-reuse-vulnerability} provide potential characterizations of code reuse gadgets and end-to-end vulnerabilities. They capture whether a value obtained through a Fetch can affect the sink and whether that Fetch is structurally related to a weakness, but do not fully enforce feasibility of the resulting end-to-end execution. A precise semantic characterization of this causal connection remains an open problem.

We also note that there are relations between the model and the identified defense strategies (\Cref{section:defenses}).
Mediation can be enforced in the \textsc{E-Store} and \textsc{E-Fetch} judgments, e.g., by requiring the index is an $n$ (rather than an $e$) when $l_o \in \sharedmem$.
Minimization amounts to reducing the size of $\sharedmem$, though it remains unclear if this can be reasonably addressed at the level of the semantics.
Constrainment can be achieved, e.g., by adding facilities to lock down objects after creation (e.g., $\kwfreeze x$) and semantics for writing to such objects.

Finally, we implement the complete $\modellanguagename$ syntax and operational semantics in an executable interpreter\footnote{Available at \url{https://github.com/KTH-LangSec/memory-safety-isnt-all-you-need}.}. The interpreter runs the model encodings in the paper and exposes shared runtime state, observations, traces, and detected weaknesses, providing an executable counterpart to the formal definitions.

\section{Related Work}\label{section:related-work}
Most work on code reuse in memory safe languages studies individual vulnerabilities, their detection, exploitation, or mitigation. In contrast, we seek a language-agnostic understanding of their common structure. We therefore discuss classical code reuse attack, code reuse attacks under memory safety, and defenses against code reuse attacks.

\tightpar{Code Reuse without Memory Safety}
Classical code reuse attacks evolved from return-to-libc to Return- and Jump-Oriented Programming~\cite{andersen2004changes,rafel1998defeating,rafel1998defeating} which rely on memory corruption. Related work also demonstrated attacks that modify data influencing control flow without directly altering control-flow targets~\cite{chen2005non,carlini2015control,acker2010protection}. These attacks target memory unsafe languages excluded by our memory safe setting.

\tightpar{Code Reuse with Memory Safety}
Prior seminal work has studied object injection~\cite{esser2010utilizing}, DOM clobbering~\cite{rydstedt2010busting,heyes2013dom}, DOM script gadgets~\cite{lekies2017code}, prototype pollution~\cite{arteau2018prototype}, and class pollution~\cite{khaled2023prototype}. Detection of these vulnerabilities have been implemented using static analysis~\cite{dahse2014code,shahriar2016object,dietrich2017evil,shcherbakov2021serialdetector,chen2024efficient,luo2024rev,li2021detecting,kim2022dapp,li2022mining,shcherbakov2023silent,liu2026first,khodayari2023dom,drescher2025dom} using taint, pattern, and graph-based approaches, dynamic analysis~\cite{koutroumpouchos2019detecting,cao2023discovering,kang2022probe,shcherbakov2024unveiling,cornelissen2024universal,khodayari2023dom,drescher2025dom} using probing, fuzzing, instrumentation, taint-aware runtimes, as well as hybrid~\cite{rasheed2020hybrid,srivastava2023crystallizer} and concolic analyses~\cite{steffens2021understanding,liu2024undefined,liu2025domino}. Beyond detection, automatic exploit generation has been implemented for object injection~\cite{dahse2014code,park2022fugio,cao2023improving}, prototype pollution~\cite{kang2025follow}, and DOM clobbering~\cite{liu2025domino}. In contrast, we abstract from attack-specific mechanisms to characterize their common structure, enabling a systematic treatment of this kind of attack, and give a formal account of the resulting weaknesses.

\tightpar{Defenses}
Prior defenses include look-ahead deserialization and execution-path restrictions for object injection~\cite{cristalli2018trusted,hawkins2017introspective,seacord2017combating}, stratification and other mitigations for prototype pollution~\cite{symbolprotoproposal,cornelissen2024universal}, countermeasures for DOM clobbering~\cite{khodayari2023dom,heiderich2017client}, and partial mitigations for DOM script gadgets~\cite{lekies2017code}. Debloating provides a more general way to reduce reusable attack surface~\cite{azad2019less,koishybayev2020mininode}. \Cref{section:defenses} relates these defenses through the framework and derives the strategies of mediation, minimization, and constraint, which enabled the identification of new defense directions.

\section{Conclusion}\label{section:conclusion}
We presented a language-agnostic framework for code reuse attacks in memory safe languages, showing how five established attack classes share a common structure based on attacker-influenced interactions with shared runtime state and subsequent reuse through gadgets. Using the framework, we derived additional weakness and gadget structures and identified three high-level defense strategies along with candidate defense directions. Finally, we developed $\modellanguagename$, a model language that gives a formal characterization of shared runtime state and code reuse attacks able to capture the established attacks. Together, these results provide a foundation for reasoning about code reuse attacks beyond individual language-specific mechanisms.

\ifcameraready%

\fi

\bibliographystyle{plain}
\bibliography{references}

@misc{pythonlanguagereference,
  author={{Python Software Foundation}},
  title={The {Python} Language Reference, 3.3: Special method names},
  howpublished={\url{https://docs.python.org/3/reference/datamodel.html#special-method-names}},
  note={Accessed 2026-10-27}
}

@online{symbolprotoproposal,
  author={Diaz, Santiago},
  title={Prototype Pollution Mitigation, {Symbol}.proto},
  year={2023},
  month={August},
  howpublished={\url{https://github.com/tc39/proposal-symbol-proto}},
  url={https://github.com/tc39/proposal-symbol-proto},
  note={Ecma TC39 Proposal, Stage 1},
  lastchecked={2026-08-20}
}

@online{heyes2013dom,
  author={Heyes, Gareth},
  title={{DOM} Clobbering},
  year={2013},
  month={May},
  howpublished={\url{https://thespanner.co.uk/2013/05/16/dom-clobbering}},
  url={https://thespanner.co.uk/2013/05/16/dom-clobbering},
  note={Blog post},
  lastchecked={2026-08-20}
}

@online{khaled2023prototype,
  author={Khaled, Abdulraheem},
  title={Prototype Pollution in {Python}},
  year={2023},
  month={January},
  howpublished={\url{https://blog.abdulrah33m.com/prototype-pollution-in-python/}},
  url={https://blog.abdulrah33m.com/prototype-pollution-in-python/},
  note={Blog post},
  lastchecked={2026-08-20}
}

@article{seacord2017combating,
  title={Combating {Java} deserialization vulnerabilities with look-ahead object input streams ({LAOIS})},
  author={Seacord, Robert},
  journal={NCC Gr Whitepaper},
  year={2017}
}

@online{rafel1998defeating,
  author={Rafel, Wojtczuk},
  title={Defeating solar designer non-executable stack patch},
  year={1998},
  howpublished={\url{https://insecure.org/sploits/non-executable.stack.problems.html}},
  url={https://insecure.org/sploits/non-executable.stack.problems.html},
  lastchecked={2026-08-20}
}

@article{andersen2004changes,
  title={Changes to functionality in {Microsoft} {Windows} {XP} service pack 2},
  author={Andersen, Starr and Abella, Vincent},
  journal={Microsoft technical document},
  howpublished={\url{https://learn.microsoft.com/en-us/previous-versions/windows/it-pro/windows-xp/bb457155(v=technet.10)?redirectedfrom=MSDN}},
  url={https://learn.microsoft.com/en-us/previous-versions/windows/it-pro/windows-xp/bb457155(v=technet.10)?redirectedfrom=MSDN},
  year={2004}
}

@article{crane2018continuing,
  title={The Continuing Arms Race: Code-Reuse Attacks and Defenses},
  author={Crane, Stephen and Homescu, Andrei and Larsen, Per and Okhravi, Hamed and Franz, Michael},
  year={2018}
}

@article{liu2025domino,
  title={The {DOM}ino Effect: Detecting and Exploiting {DOM} Clobbering Gadgets via Concolic Execution with Symbolic {DOM}},
  author={Liu, Zhengyu and Lee, Theo and Yu, Jianjia and Kang, Zifeng and Cao, Yinzhi}
}

@article{rydstedt2010busting,
  title={Busting frame busting: A study of clickjacking vulnerabilities at popular sites},
  author={Rydstedt, Gustav and Bursztein, Elie and Boneh, Dan and Jackson, Collin},
  journal={IEEE Oakland Web},
  volume={2},
  number={6},
  pages={24},
  year={2010}
}

@article{esser2010utilizing,
  title={Utilizing code reuse/{ROP} in {PHP} application exploits},
  author={Esser, Stefan},
  journal={Proceedings of the Black Hat USA},
  year={2010}
}

@article{kim2022dapp,
  title={{DAPP}: Automatic detection and analysis of prototype pollution vulnerability in {Node.js} modules},
  author={Kim, Hee Yeon and Kim, Ji Hoon and Oh, Ho Kyun and Lee, Beom Jin and Mun, Si Woo and Shin, Jeong Hoon and Kim, Kyounggon},
  journal={International Journal of Information Security},
  volume={21},
  number={1},
  pages={1--23},
  year={2022},
  publisher={Springer}
}

@article{bracha2004mirrors,
  title={Mirrors: Design principles for meta-level facilities of object-oriented programming languages},
  author={Bracha, Gilad and Ungar, David},
  journal={ACM SIGPLAN Notices},
  volume={39},
  number={10},
  pages={331--344},
  year={2004},
  publisher={ACM New York, NY, USA}
}

@article{steffens2021understanding,
  title={Understanding emerging client-side web vulnerabilities using dynamic program analysis},
  author={Steffens, Marius},
  year={2021},
}

@inproceedings{chen2024efficient,
  title={Efficient detection of {Java} deserialization gadget chains via bottom-up gadget search and dataflow-aided payload construction},
  author={Chen, Bofei and Zhang, Lei and Huang, Xinyou and Cao, Yinzhi and Lian, Keke and Zhang, Yuan and Yang, Min},
  booktitle={2024 IEEE Symposium on Security and Privacy (SP)},
  pages={3961--3978},
  year={2024},
  organization={IEEE}
}

@inproceedings{cornelissen2024universal,
  title={{GHunter}: Universal Prototype Pollution Gadgets in {JavaScript} Runtimes},
  author={Cornelissen, Eric and Shcherbakov, Mikhail and Balliu, Musard},
  booktitle={33rd USENIX Security Symposium (USENIX Security 24)},
  pages={3693--3710},
  year={2024}
}

@inproceedings{kang2022probe,
  title={Probe the Proto: Measuring Client-Side Prototype Pollution Vulnerabilities of One Million Real-world Websites.},
  author={Kang, Zifeng and Li, Song and Cao, Yinzhi},
  booktitle={NDSS},
  year={2022}
}

@inproceedings{kang2025follow,
  title={Follow My Flow: Unveiling Client-Side Prototype Pollution Gadgets from One Million Real-World Websites},
  author={Kang, Zifeng and Lyu, Muxi and Liu, Zhengyu and Yu, Jianjia and Fan, Runqi and Li, Song and Cao, Yinzhi},
  booktitle={2025 IEEE Symposium on Security and Privacy (SP)},
  pages={991--1008},
  year={2025},
  organization={IEEE}
}

@inproceedings{lekies2017code,
  title={Code-reuse attacks for the web: Breaking cross-site scripting mitigations via script gadgets},
  author={Lekies, Sebastian and Kotowicz, Krzysztof and Gro{\ss}, Samuel and Vela Nava, Eduardo A and Johns, Martin},
  booktitle={Proceedings of the 2017 ACM SIGSAC Conference on Computer and Communications Security},
  pages={1709--1723},
  year={2017}
}

@inproceedings{liu2024undefined,
  title={Undefined-oriented programming: Detecting and chaining prototype pollution gadgets in {Node.js} template engines for malicious consequences},
  author={Liu, Zhengyu and An, Kecheng and Cao, Yinzhi},
  booktitle={2024 IEEE Symposium on Security and Privacy (SP)},
  pages={4015--4033},
  year={2024},
  organization={IEEE}
}

@inproceedings{shcherbakov2021serialdetector,
  title={{SerialDetector}: Principled and Practical Exploration of Object Injection Vulnerabilities for the Web},
  author={Shcherbakov, Mikhail and Balliu, Musard},
  booktitle={Network and Distributed Systems Security (NDSS) Symposium 2021-21-24 February 2021},
  year={2021}
}

@inproceedings{shcherbakov2023silent,
  title={Silent Spring: Prototype Pollution Leads to Remote Code Execution in {Node.js}},
  author={Shcherbakov, Mikhail and Balliu, Musard and Staicu, Cristian-Alexandru},
  booktitle={32nd USENIX Security Symposium (USENIX Security 23)},
  pages={5521--5538},
  year={2023}
}

@inproceedings{shcherbakov2024unveiling,
  title={Unveiling the invisible: Detection and evaluation of prototype pollution gadgets with dynamic taint analysis},
  author={Shcherbakov, Mikhail and Moosbrugger, Paul and Balliu, Musard},
  booktitle={Proceedings of the ACM Web Conference 2024},
  pages={1800--1811},
  year={2024}
}

@inproceedings{khodayari2023dom,
  title={It's ({DOM}) clobbering time: Attack techniques, prevalence, and defenses},
  author={Khodayari, Soheil and Pellegrino, Giancarlo},
  booktitle={2023 IEEE Symposium on Security and Privacy (SP)},
  pages={1041--1058},
  year={2023},
  organization={IEEE}
}

@inproceedings{li2021detecting,
  title={Detecting {Node.js} prototype pollution vulnerabilities via object lookup analysis},
  author={Li, Song and Kang, Mingqing and Hou, Jianwei and Cao, Yinzhi},
  booktitle={Proceedings of the 29th ACM Joint Meeting on European Software Engineering Conference and Symposium on the Foundations of Software Engineering},
  pages={268--279},
  year={2021}
}

@inproceedings{li2022mining,
  title={Mining {Node.js} vulnerabilities via object dependence graph and query},
  author={Li, Song and Kang, Mingqing and Hou, Jianwei and Cao, Yinzhi},
  booktitle={31st USENIX Security Symposium (USENIX Security 22)},
  pages={143--160},
  year={2022}
}

@inproceedings{heiderich2017client,
  title={{DOMPurify}: Client-side protection against {XSS} and markup injection},
  author={Heiderich, Mario and Sp{\"a}th, Christopher and Schwenk, J{\"o}rg},
  booktitle={European Symposium on Research in Computer Security},
  pages={116--134},
  year={2017},
  organization={Springer}
}

@inproceedings{dahse2014code,
  title={Code reuse attacks in {PHP}: Automated {POP} chain generation},
  author={Dahse, Johannes and Krein, Nikolai and Holz, Thorsten},
  booktitle={Proceedings of the 2014 ACM SIGSAC Conference on Computer and Communications Security},
  pages={42--53},
  year={2014}
}

@inproceedings{park2022fugio,
  title={{FUGIO}: Automatic exploit generation for {PHP} object injection vulnerabilities},
  author={Park, Sunnyeo and Kim, Daejun and Jana, Suman and Son, Sooel},
  booktitle={31st USENIX Security Symposium (USENIX Security 22)},
  pages={197--214},
  year={2022}
}

@inproceedings{drescher2025dom,
  title={In the {DOM} We Trust: Exploring the Hidden Dangers of Reading from the {DOM} on the Web},
  author={Drescher, Jan and Mirzaei, Sepehr and Khodayari, Soheil and Klein, David and Barber, Thomas and Johns, Martin and Pellegrino, Giancarlo},
  booktitle={Proceedings of the 2025 ACM SIGSAC Conference on Computer and Communications Security},
  year={2025}
}

@inproceedings{liu2026first,
  title={The First Large-Scale Systematic Study of {Python} Class Pollution Vulnerability},
  author={Liu, Zhengyu and Zhong, Jiacheng and Yu, Jianjia and Lyu, Muxi and Kang, Zifeng and Cao, Yinzhi},
  booktitle={2026 IEEE Symposium on Security and Privacy (SP)},
  year={2026}
}

@inproceedings{koutroumpouchos2019detecting,
  title={{ObjectMap}: Detecting insecure object deserialization},
  author={Koutroumpouchos, Nikolaos and Lavdanis, Georgios and Veroni, Eleni and Ntantogian, Christoforos and Xenakis, Christos},
  booktitle={Proceedings of the 23rd Pan-Hellenic Conference on Informatics},
  pages={67--72},
  year={2019}
}

@inproceedings{koishybayev2020mininode,
  title={Mininode: Reducing the attack surface of {Node.js} applications},
  author={Koishybayev, Igibek and Kapravelos, Alexandros},
  booktitle={23rd international symposium on research in attacks, intrusions and defenses (RAID 2020)},
  pages={121--134},
  year={2020}
}

@inproceedings{azad2019less,
  title={Less is more: Quantifying the security benefits of debloating web applications},
  author={Azad, Babak Amin and Laperdrix, Pierre and Nikiforakis, Nick},
  booktitle={28th USENIX Security Symposium (USENIX Security 19)},
  pages={1697--1714},
  year={2019}
}

@inproceedings{chen2005non,
  title={Non-control-data attacks are realistic threats.},
  author={Chen, Shuo and Xu, Jun and Sezer, Emre Can and Gauriar, Prachi and Iyer, Ravishankar K},
  booktitle={USENIX security symposium},
  volume={5},
  pages={146},
  year={2005}
}

@inproceedings{carlini2015control,
  title={{Control-Flow} bending: On the effectiveness of Control-Flow integrity},
  author={Carlini, Nicholas and Barresi, Antonio and Payer, Mathias and Wagner, David and Gross, Thomas R},
  booktitle={24th USENIX Security Symposium (USENIX Security 15)},
  pages={161--176},
  year={2015}
}

@inproceedings{shahriar2016object,
  title={Object injection vulnerability discovery based on latent semantic indexing},
  author={Shahriar, Hossain and Haddad, Hisham},
  booktitle={Proceedings of the 31st Annual ACM Symposium on Applied Computing},
  pages={801--807},
  year={2016}
}

@inproceedings{cristalli2018trusted,
  title={Trusted execution path for protecting {Java} applications against deserialization of untrusted data},
  author={Cristalli, Stefano and Vignati, Edoardo and Bruschi, Danilo and Lanzi, Andrea},
  booktitle={International Symposium on Research in Attacks, Intrusions, and Defenses},
  pages={445--464},
  year={2018},
  organization={Springer}
}

@inproceedings{hawkins2017introspective,
  title={{ZenIDS}: Introspective intrusion detection for {PHP} applications},
  author={Hawkins, Byron and Demsky, Brian},
  booktitle={2017 IEEE/ACM 39th International Conference on Software Engineering (ICSE)},
  pages={232--243},
  year={2017},
  organization={IEEE}
}

@inproceedings{dietrich2017evil,
  title={Evil pickles: {DoS} attacks based on object-graph engineering},
  author={Dietrich, Jens and Jezek, Kamil and Rasheed, Shawn and Tahir, Amjed and Potanin, Alex},
  booktitle={31st European Conference on Object-Oriented Programming (ECOOP 2017)},
  pages={10--1},
  year={2017},
  organization={Schloss Dagstuhl--Leibniz-Zentrum f{\"u}r Informatik}
}

@inproceedings{cao2023improving,
  title={Improving {Java} deserialization gadget chain mining via overriding-guided object generation},
  author={Cao, Sicong and Sun, Xiaobing and Wu, Xiaoxue and Bo, Lili and Li, Bin and Wu, Rongxin and Liu, Wei and He, Biao and Ouyang, Yu and Li, Jiajia},
  booktitle={2023 IEEE/ACM 45th International Conference on Software Engineering (ICSE)},
  pages={397--409},
  year={2023},
  organization={IEEE}
}

@inproceedings{cao2023discovering,
  title={{ODDFuzz}: Discovering {Java} deserialization vulnerabilities via structure-aware directed greybox fuzzing},
  author={Cao, Sicong and He, Biao and Sun, Xiaobing and Ouyang, Yu and Zhang, Chao and Wu, Xiaoxue and Su, Ting and Bo, Lili and Li, Bin and Ma, Chuanlei},
  booktitle={2023 IEEE Symposium on Security and Privacy (SP)},
  pages={2726--2743},
  year={2023},
  organization={IEEE}
}

@inproceedings{luo2024rev,
  title={Rev Gadget: A {Java} Deserialization Gadget Chains Discover Tool Based on Reverse Semantics and Taint Analysis},
  author={Luo, Yifan and Cui, Baojiang},
  booktitle={International Conference on Emerging Internet, Data \& Web Technologies},
  pages={229--240},
  year={2024},
  organization={Springer}
}

@inproceedings{acker2010protection,
  title={{ValueGuard}: Protection of native applications against data-only buffer overflows},
  author={Van Acker, Steven and Nikiforakis, Nick and Philippaerts, Pieter and Younan, Yves and Piessens, Frank},
  booktitle={International Conference on Information Systems Security},
  pages={156--170},
  year={2010},
  organization={Springer}
}

@inproceedings{rasheed2020hybrid,
  title={A hybrid analysis to detect java serialisation vulnerabilities},
  author={Rasheed, Shawn and Dietrich, Jens},
  booktitle={Proceedings of the 35th IEEE/ACM International Conference on Automated Software Engineering},
  pages={1209--1213},
  year={2020}
}

@inproceedings{srivastava2023crystallizer,
  title={Crystallizer: A hybrid path analysis framework to aid in uncovering deserialization vulnerabilities},
  author={Srivastava, Prashast and Toffalini, Flavio and Vorobyov, Kostyantyn and Gauthier, Fran{\c{c}}ois and Bianchi, Antonio and Payer, Mathias},
  booktitle={Proceedings of the 31st ACM Joint European Software Engineering Conference and Symposium on the Foundations of Software Engineering},
  pages={1586--1597},
  year={2023}
}

@inproceedings{janc2020oh,
  title={Oh, the places you'll go! finding our way back from the web platform's ill-conceived jaunts},
  author={Janc, Artur and West, Mike},
  booktitle={2020 IEEE European Symposium on Security and Privacy Workshops (EuroS\&PW)},
  pages={673--680},
  year={2020},
  organization={IEEE}
}

@incollection{arteau2018prototype,
  title={Prototype pollution attack in {NodeJS} application},
  author={Arteau, Olivier},
  booktitle={NorthSec},
  year={2018}
}

\appendix

\section{Open Science}\label{section:open-science}
We provide runnable versions of \Cref{listing:object-injection}--\ref{listing:horizontal-data-dependency}.
We also provide an interpreter for $\modellanguagename$ as described in \Cref{section:model} along with runnable versions of all listings in the paper as well as all canonical weakness patterns (corresponding to paths in \Cref{figure:categorization:weaknesses}) and gadget patterns (corresponding to rows in \Cref{table:code-reuse-gadgets}) from \Cref{section:categorization}. The interpreter implements the full syntax and operational semantics, including observations and traces, and pretty printing of intermediate configurations and final states.
These are available at:

\noindent\url{https://github.com/KTH-LangSec/memory-safety-isnt-all-you-need}

\begin{figure*}[t]
  \centering
  \begin{mathpar}
      \frac{
        -
      }{
        \langle \kwskip, \Store, \Heap, \Functions, \Context \rangle
          \step
        \langle \epsilon, \Store, \Heap, \Functions, \Context \rangle
      } \; \text{(\textsc{E-Skip})}
      \quad
      \frac{
        \langle c_1, \Store, \Heap, \Functions, \Context \rangle
          \stackrel{\observation}{\step}
        \langle c_1', \Store', \Heap', \Functions', \Context' \rangle
      }{
        \langle c_1; c_2, \Store, \Heap, \Functions, \Context \rangle
          \stackrel{\observation}{\step}
        \langle c_1'; c_2, \Store', \Heap', \Functions', \Context' \rangle
      } \; \text{(\textsc{E-Seq})}

      \frac{
        \langle c_1, \Store, \Heap, \Functions, \Context \rangle
          \stackrel{\observation}{\step}
        \langle \epsilon, \Store', \Heap', \Functions', \Context' \rangle
      }{
        \langle c_1; c_2, \Store, \Heap, \Functions, \Context \rangle
          \stackrel{\observation}{\step}
        \langle c_2, \Store', \Heap', \Functions', \Context' \rangle
      } \; \text{(\textsc{E-Seq-}$\epsilon$)}
      \quad
      \frac{
        v = [\![e]\!]_{\State}
      }{
        \langle x := e, \Store, \Heap, \Functions, \Context \rangle
          \step
        \langle \epsilon, \Store[x \mapsto v], \Heap, \Functions, \Context \rangle
      } \; \text{(\textsc{E-Assign})}
      \quad

      \frac{
        l_f \not\in \Functions
      }{
        \langle \kwfn x \kwblock{c}, \Store, \Heap, \Functions, \Context \rangle
          \step
        \langle \epsilon, \Store[x \mapsto l_f], \Heap, \Functions[l_f \mapsto c], \Context \rangle
      } \; \text{(\textsc{E-Func})}
      \quad
      \frac{
        l_f = \State(x)
          \quad
        l_f \in \locations_{\Functions}
          \quad
        c = \Functions(l_f)
      }{
        \langle \kwcall x, \Store, \Heap, \Functions, \Context \rangle
          \step
        \langle c, \Store, \Heap, \Functions, \Context \rangle
      } \; \text{(\textsc{E-Call})}

      \frac{
        n = [\![e]\!]_{\State}
          \quad
        n \in \mathbb{Z}
          \quad
        n \neq 0
      }{
        \langle \kwif e \kwblock{c_1} \kwelse \kwblock{c_2}, \Store, \Heap, \Functions, \Context \rangle
          \step
        \langle c_1, \Store, \Heap, \Functions, \Context \rangle
      } \; \text{(\textsc{E-If-Then})}
      \quad
      \frac{
        [\![e]\!]_{\State} = 0
      }{
        \langle \kwif e \kwblock{c_1} \kwelse \kwblock{c_2}, \Store, \Heap, \Functions, \Context \rangle
          \step
        \langle c_2, \Store, \Heap, \Functions, \Context \rangle
      } \; \text{(\textsc{E-If-Else})}

      \frac{
        -
      }{
        \langle \kwwhile e \kwblock{c}, \Store, \Heap, \Functions, \Context \rangle
          \step
        \langle \kwif e \kwblock{c; \; \kwwhile e \kwblock{c}} \kwelse \kwblock{\kwskip}, \Store, \Heap, \Functions, \Context \rangle
      } \; \text{(\textsc{E-While})}
  \end{mathpar}
  \caption{
    The operation semantics for the remaining commands not covered in \Cref{figure:semantics}.
  }\label{figure:semantics-rest}
\end{figure*}

\section{Categorization, Extended}\label{section:categorization-extended}
This section expands on \Cref{section:categorization}.

\subsection{Weakness Cycles}
The framework (\Cref{figure:framework}) permits cycling through \texttt{Store} and \texttt{Fetch} operations. We are not aware of any real-world examples of such a weakness, yet our framework supports describing scenarios requiring them. Such scenarios seem to correspond to cases where either $l_s$, $v$, or $l_f$ is constrained in some way.

An example scenario is a case where it is possible to mutate one area of shared runtime state and recycle between that and another area of shared runtime state ($l_s$ is constrained). Together these allow mutating the second area of shared memory through the path
\begin{equation*}
  \begin{aligned}
    \text{Data} & \xrightarrow{\texttt{store}(l_{a1}, d)} \text{Store}
                  \xrightarrow{\texttt{fetch}(l_{a1}')} \text{Fetch} & \textit{(mutate)} \\
                & \xrightarrow{\texttt{store}(l_{a2}, d')} \text{Store}
                  \xrightarrow{\texttt{fetch}(l_{a2}')} \text{Fetch} & \textit{(recycle)} \\
  \end{aligned}
\end{equation*}

\subsection{Control}
\begin{figure*}[]
  \centering
  \begin{subfigure}[b]{0.18\linewidth}
    \centering
    \includegraphics[width=\linewidth]{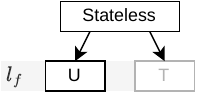}
    \caption{}\label{figure:categorization:stateless-weakness}
  \end{subfigure}
  \hfill
  \begin{subfigure}[b]{0.39\linewidth}
    \centering
    \includegraphics[width=\linewidth]{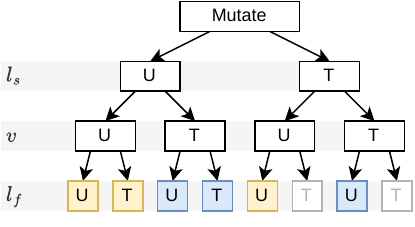}
    \caption{}\label{figure:categorization:mutating-weakness}
  \end{subfigure}
  \hfill
  \begin{subfigure}[b]{0.39\linewidth}
    \centering
    \includegraphics[width=\linewidth]{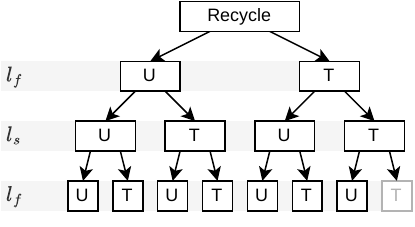}
    \caption{}\label{figure:categorization:recycling-weakness}
  \end{subfigure}
  \caption{
    Categorization of stateless (\ref{figure:categorization:stateless-weakness}), mutating (\ref{figure:categorization:mutating-weakness}), and recycling (\ref{figure:categorization:recycling-weakness}) code reuse attacks.
    `T' stands for trusted and `U' for untrusted.
    Grayed out leafs are not code reuse and the highlights denote \colorbox[HTML]{FFF2CC}{payload} and \colorbox[HTML]{DAE8FC}{fixed} weaknesses.
  }\label{figure:categorization:weaknesses}
\end{figure*}

\Cref{figure:categorization:weaknesses} shows a full categorization of all possible stateless, mutating, and recycling weaknesses corresponding to the parameters of \texttt{fetch($l_f$)} and \texttt{store($l_s$, $v$)}. The stateful categorization is trivial and the mutating categorization is discussed in \Cref{section:categorization}. While the recycling categorization is not trivial, we did not gain additional insight from it.
One observation worth noting is that the subtree at $\text{T}\to\text{T}$ in the mutating (and recycling) categorization corresponds to the stateless case  (i.e., a stateless attack requires that a trusted value was stored at a trusted location).

\subsection{Gadget}
\begin{figure*}[]
  \centering
  \includegraphics[width=0.45\linewidth]{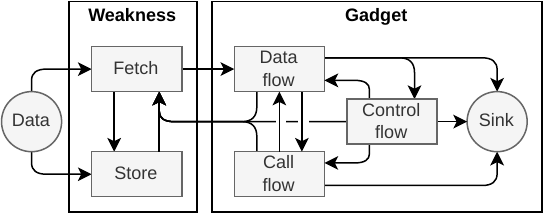}
  \caption{
    Extended framework of code reuse attacks.
    Extends \Cref{figure:framework}.
  }\label{figure:framework-extended}
\end{figure*}

\begin{figure*}[]
  \centering
  \begin{subfigure}[b]{0.49\linewidth}
    \centering
    \includegraphics[width=\linewidth]{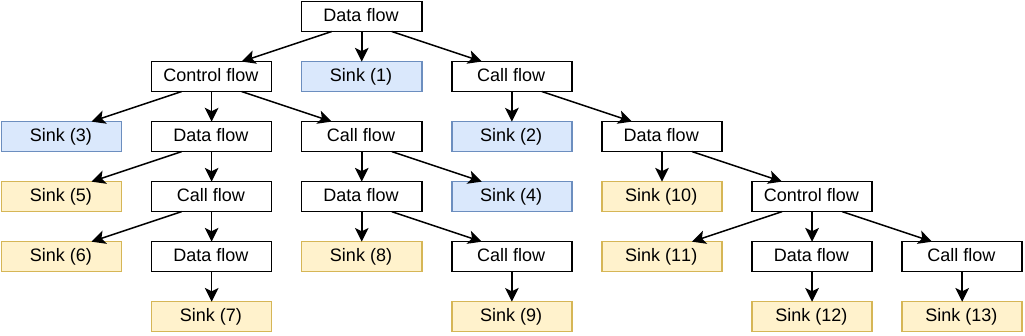}
    \caption{}\label{figure:categorization:simple-and-vertical-gadgets}
  \end{subfigure}
  \hfill
  \begin{subfigure}[b]{0.49\linewidth}
    \centering
    \includegraphics[width=\linewidth]{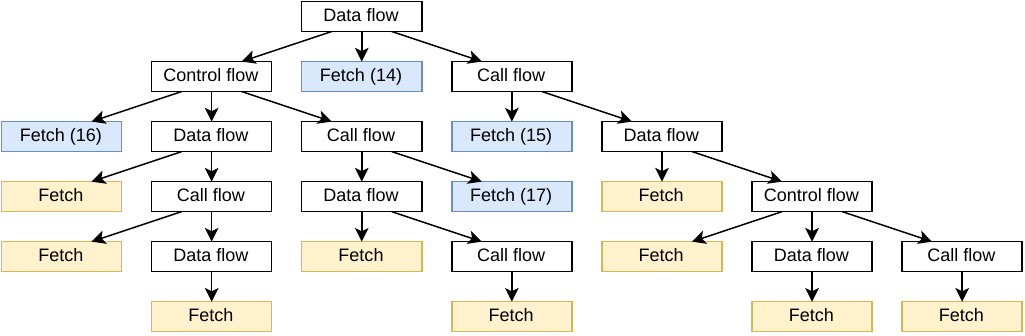}
    \caption{
    }\label{figure:categorization:horizontal-gadgets}
  \end{subfigure}
  \caption{
    Categorization of \colorbox[HTML]{DAE8FC}{simple} and \colorbox[HTML]{FFF2CC}{vertical} gadgets (\ref{figure:categorization:simple-and-vertical-gadgets}) and horizontal gadgets, \colorbox[HTML]{DAE8FC}{minimal} and \colorbox[HTML]{FFF2CC}{vertical prefix} bearing (\ref{figure:categorization:horizontal-gadgets}).
    Leaf numbering corresponds to \Cref{table:code-reuse-gadgets}.
  }\label{figure:categorization:gadgets}
\end{figure*}

\Cref{figure:framework-extended} presents the extended version of our framework (\Cref{figure:framework}) as described in \Cref{section:categorization:gadget}.
\Cref{figure:categorization:gadgets} shows a full categorization of all minimal gadget, covering simple gadgets, vertical gadget chains, and horizontal gadget chains.
\Cref{table:code-reuse-gadgets} provides an overview of minimal code snippets corresponding to the minimal gadgets

\section{Model, Extended}\label{section:model-extended}

This section completes the operation semantics, see \Cref{figure:semantics-rest}, as well as the modeling of known attacks started in \Cref{section:model:known-attacks}.

\subsection{Class Pollution}\label{section:model:class-pollution}
\Cref{listing:model:class-pollution} encodes \Cref{listing:class-pollution} in $\modellanguagename$.
Recall that this code allows the attacker to alter a static property of class \texttt{B} starting from an instance of class \texttt{A} through careful navigation of the object graph using dynamic property lookups and assignment. We model the object graph by setting up a series of references (line 3--9). Then, the property lookups and assignment are performed (line 11) and finally \texttt{B}'s property is accessed (line 13) and outputted (line 15).

\begin{figure}[]
  \centering
  \begin{equation*}
    \begin{aligned}
       1 \quad & \texttt{\_\_globals\_\_} := \texttt{0}; \; \texttt{\_\_class\_\_} := \texttt{1}; \; \texttt{\_\_init\_\_} := \texttt{2};                                                                      \\
         \quad & \quad \quad \texttt{classA} := \texttt{3}; \; \texttt{classB} := \texttt{4}; \; \texttt{p} := \texttt{5};                                                                                     \\
       2 \quad & \texttt{\textcolor{gray}{// Initialization}}                                                                                                                                                  \\
       3 \quad & \texttt{global} \gets \kwalloc_{\Context};                                                                                                                                                    \\
       4 \quad & \texttt{A} \gets \kwalloc; \; \texttt{global}[\texttt{classA}] := \texttt{A}; \; \texttt{init} \gets \kwalloc;                                                                                \\
       5 \quad & \texttt{A}[\texttt{\_\_init\_\_}] := \texttt{init}; \; \texttt{init}[\texttt{\_\_globals\_\_}] := \texttt{global};                                                                            \\
       6 \quad & \texttt{B} \gets \kwalloc; \; \texttt{global}[\texttt{classB}] := \texttt{B}; \; \texttt{init} \gets \kwalloc;                                                                                \\
       7 \quad & \texttt{B}[\texttt{\_\_init\_\_}] := \texttt{init}; \; \texttt{init}[\texttt{\_\_globals\_\_}] := \texttt{global};                                                                            \\
       8 \quad & \texttt{B}[\texttt{p}] := \texttt{42};                                                                                                                                                        \\
       9 \quad & \texttt{o} \gets \kwalloc; \; \texttt{o}[\texttt{\_\_class\_\_}] := \texttt{A};                                                                                                               \\
      10 \quad & \texttt{\textcolor{gray}{// `setattr(getattr(...)[i4], i5, i6)'}}                                                                                                                             \\
      11 \quad & \texttt{t} := \texttt{o}[\Input_1]; \; \texttt{t} := \texttt{t}[\Input_2]; \; \texttt{t} := \texttt{t}[\Input_3]; \; \texttt{t} := \texttt{t}[\Input_4]; \; \texttt{t}[\Input_5] := \Input_6; \\
      12 \quad & \texttt{\textcolor{gray}{// `x = B.p'}}                                                                                                                                                       \\
      13 \quad & \texttt{x} := \texttt{B}[\texttt{p}];                                                                                                                                                         \\
      14 \quad & \texttt{\textcolor{gray}{// `eval(x)'}}                                                                                                                                                       \\
      15 \quad & \Output := \texttt{x}                                                                                                                                                                         \\
    \end{aligned}
  \end{equation*}
  \captionsetup{type=lstlisting}
  \caption{
    Class pollution in $\modellanguagename$.
  }\label{listing:model:class-pollution}
\end{figure}

For this program we can demonstrate how the security conditions capture its exploitation. First, the input pair
\begin{equation*}
  \begin{aligned}
    &\Store_1[\Input_1 \mapsto 1, \Input_2 \mapsto 2, \Input_3 \mapsto 0, \Input_4 \mapsto 4, \Input_5 \mapsto 6, \Input_6 \mapsto 7]  \\
    &\Store_2[\Input_1 \mapsto 1, \Input_2 \mapsto 2, \Input_3 \mapsto 0, \Input_4 \mapsto 4, \Input_5 \mapsto 5, \Input_6 \mapsto 67] \\
  \end{aligned}
\end{equation*}
produces the traces (identical write observations omitted for brevity)
\begin{equation*}
  \begin{aligned}
    \trace_1 &= \multiset{\cdots, \observation^w_{[\![\texttt{B}]\!],6}} \\
    \trace_2 &= \multiset{\cdots, \observation^w_{[\![\texttt{B}]\!],5}} \\
  \end{aligned}
\end{equation*}
given $\trustedmem = \emptyset$ we have $\sharedmem^u_i = \{ [\![\texttt{global}]\!], \cdots, [\![\texttt{B}]\!] \}$ and find
\begin{equation*}
  \begin{aligned}
                                \trace_1 \cap \sharedmem^u_1 &\neq \trace_2 \cap \sharedmem^u_2                             \\
    \multiset{\cdots, \observation^w_{[\![\texttt{B}]\!],6}} &\neq \multiset{\cdots, \observation^w_{[\![\texttt{B}]\!],5}} \\
  \end{aligned}
\end{equation*}
satisfying the security condition for a code reuse weakness.
If we consider the second execution, we can execute until $\observation^r_{[\![\texttt{B}]\!],5}$ (corresponding to $\texttt{B}[\texttt{p}]$, line 13) and continue execution with $v_1=6$ and $v_2=7$ until completion to observe different outputs, $\texttt{6}$ and $\texttt{7}$ resp., thus satisfying the security condition for a code reuse gadget.
Since we used an initial state of the weakness to demonstrate the gadget, and the modified observation is related to an observation in the delta between the two executions prior to it, namely $\observation^w_{[\![\texttt{B}]\!],5}$, the security condition for a code reuse vulnerability is also met.
Moreover, because $\trace^w_1 \cap \sharedmem^u_1 \neq \trace^w_2 \cap \sharedmem^u_2$ we know this is a \emph{stateful} vulnerability.

\subsection{DOM Clobbering}\label{section:model:dom-clobbering}
\Cref{listing:model:dom-clobbering} encodes \Cref{listing:dom-clobbering} in $\modellanguagename$.
Recall that this attack is possible because the attacker controls the attributes, including a \texttt{name} (or \texttt{id}), of an element in the DOM. We model the DOM as a flat list of elements stored in \texttt{body} (line 3--6). The attacker's element is created (line 7) and inserted into the DOM (line 9) after which a named property lookup of \texttt{currentScript} is performed on the \texttt{document} (line 11--15). Finally, the \texttt{src} of the \texttt{currentScript} is accessed and outputted (line 17).

\begin{figure}[]
  \centering
  \begin{equation*}
    \begin{aligned}
       1 \quad & \texttt{len} := 0; \; \texttt{DOM} := \texttt{1}; \; \texttt{currentScript} := \texttt{2}; \; \texttt{name} := \texttt{3};                    \\
         \quad & \quad \quad \texttt{src} := \texttt{4};                                                                                                       \\
       2 \quad & \texttt{\textcolor{gray}{// Initialization}}                                                                                                  \\
       3 \quad & \texttt{body} \gets \kwalloc; \; \texttt{body}[\texttt{len}] := \texttt{0};                                                                   \\
       4 \quad & \texttt{cur} \gets \kwalloc; \; \texttt{cur}[\texttt{src}] := \texttt{42};                                                                    \\
       5 \quad & \texttt{document} \gets \kwalloc_{\Context}; \; \texttt{document}[\texttt{DOM}] := \texttt{body};                                             \\
       6 \quad & \texttt{document}[\texttt{currentScript}] := \texttt{cur};                                                                                    \\
       7 \quad & \texttt{el} \gets \kwalloc; \; \texttt{el}[\Input_1] := \Input_2; \; \texttt{el}[\Input_3] := \Input_4;                                       \\
       8 \quad & \texttt{\textcolor{gray}{// `document.body.appendChild(el)'}}                                                                                 \\
       9 \quad & \texttt{l} := \texttt{body}[\texttt{len}]; \; \texttt{body}[\texttt{l + 1}] := \texttt{el}; \; \texttt{body}[\texttt{len}] := \texttt{l + 1}; \\
      10 \quad & \texttt{\textcolor{gray}{// `var x = document.currentScript'}}                                                                                \\
      11 \quad & \texttt{x} := \nil; \; \texttt{p} := \texttt{currentScript}; \; \texttt{i} := \texttt{0}; \; \texttt{l} := \texttt{body}[\texttt{len}];       \\
      12 \quad & \kwwhile (\texttt{x} = \nil) \land (\texttt{i} \neq \texttt{l}) \;\{                                                                          \\
      13 \quad & \quad \texttt{i} := \texttt{i + 1}; \; \texttt{c} := \texttt{body}[\texttt{i}]; \; \texttt{n} := \texttt{c}[\texttt{name}];                   \\
      14 \quad & \quad \kwif \texttt{n} = \texttt{p} \kwblock{\texttt{x} := \texttt{c}} \kwelse \kwblock{\kwskip} \;\}                                         \\
      15 \quad & \kwif \texttt{x} = \nil \kwblock{\texttt{x} := \texttt{document}[\texttt{p}]} \kwelse \kwblock{\kwskip}                                       \\
      16 \quad & \texttt{\textcolor{gray}{// `script.src = x.src'}}                                                                                            \\
      17 \quad & \Output := \texttt{x}[\texttt{src}]                                                                                                           \\
    \end{aligned}
  \end{equation*}
  \captionsetup{type=lstlisting}
  \caption{
    DOM clobbering in $\modellanguagename$.
  }\label{listing:model:dom-clobbering}
\end{figure}

For this program we can demonstrate how the security conditions capture its exploitation. First, the input pair
\begin{equation*}
  \begin{aligned}
    &\Store_1[\Input_1 \mapsto 1, \Input_2 \mapsto 2, \Input_3 \mapsto 3, \Input_4 \mapsto 4]  \\
    &\Store_2[\Input_1 \mapsto 3, \Input_2 \mapsto 2, \Input_3 \mapsto 4, \Input_4 \mapsto 67] \\
  \end{aligned}
\end{equation*}
produces the traces (identical write observations omitted for brevity)
\begin{equation*}
  \begin{aligned}
    \trace_1 &= \multiset{\cdots, \observation^w_{[\![\texttt{el}]\!],1}, \observation^r_{[\![\texttt{document}]\!],2}, \observation^r_{[\![\texttt{cur}]\!],4}} \\
    \trace_2 &= \multiset{\cdots, \observation^w_{[\![\texttt{el}]\!],4}, \observation^r_{[\![\texttt{el}]\!],4}}                                                \\
  \end{aligned}
\end{equation*}
given $\trustedmem = \emptyset$ we have $\sharedmem^u_i = \{ [\![\texttt{body}]\!], \cdots, [\![\texttt{el}]\!] \}$ and find
\begin{equation*}
  \begin{aligned}
                                                                                                                        \trace_1 \cap \sharedmem^u_1 &\neq \trace_2 \cap \sharedmem^u_2                                                                      \\
    \multiset{\cdots, \observation^w_{[\![\texttt{el}]\!],1}, \observation^r_{[\![\texttt{document}]\!],2}, \observation^r_{[\![\texttt{cur}]\!],4}} &\neq \multiset{\cdots, \observation^w_{[\![\texttt{el}]\!],4}, \observation^r_{[\![\texttt{el}]\!],4}} \\
  \end{aligned}
\end{equation*}
violating the security condition for a code reuse weakness.
If we consider the second execution, we can execute until $\observation^r_{[\![\texttt{el}]\!],4}$ (corresponding to $\texttt{x}[\texttt{src}]$, line 17) and continue execution with $v_1=6$ and $v_2=7$ until completion to observe different outputs, $\texttt{6}$ and $\texttt{7}$ resp., thus satisfying the security condition for a code reuse gadget.
Since we used an initial state of the weakness to demonstrate the gadget, and the modified observation is related to an observation in the delta between the two executions prior to it, namely $\observation^w_{[\![\texttt{el}]\!],4}$, the security condition for a code reuse vulnerability is also met.
Moreover, because $\trace^w_1 \cap \sharedmem^u_1 \neq \trace^w_2 \cap \sharedmem^u_2$ we know this is a \emph{stateful} vulnerability.

\subsection{DOM Script Gadgets}\label{section:model:dom-script-gadgets}
\Cref{listing:model:dom-script-gadgets} encodes \Cref{listing:dom-script-gadgets} in $\modellanguagename$.
Recall that this attack works by controlling part of the DOM. We use the same model of the DOM as before (line 3--4). The attacker's element is created (line 5) and inserted into the DOM (line 7) after which the code looks for an element with id \texttt{button} (line 9--12) for which it gets the \texttt{dataText} value and outputs it (line 14).

\begin{figure}[]
  \centering
  \begin{equation*}
    \begin{aligned}
       1 \quad & \texttt{len} := \texttt{0}; \; \texttt{DOM} := \texttt{1}; \; \texttt{id} := \texttt{2}; \; \texttt{dataText} := \texttt{3};                \\
         \quad & \quad \quad \texttt{button} := \texttt{4};                                                                                                  \\
       2 \quad & \texttt{\textcolor{gray}{// Initialization}}                                                                                                \\
       3 \quad & \texttt{body} \gets \kwalloc; \; \texttt{body}[\texttt{len}] := \texttt{0};                                                                 \\
       4 \quad & \texttt{document} \gets \kwalloc_{\Context}; \; \texttt{document}[\texttt{DOM}] := \texttt{body};                                           \\
       5 \quad & \texttt{el} \gets \kwalloc; \; \texttt{el}[\Input_1] := \Input_2; \; \texttt{el}[\Input_3] := \Input_4;                                     \\
       6 \quad & \texttt{\textcolor{gray}{// `document.body.appendChild(el)'}}                                                                               \\
       7 \quad & \texttt{l} := \texttt{body}[\texttt{len}]; \; \text{body}[\texttt{l + 1}] := \texttt{el}; \; \texttt{body}[\texttt{len}] := \texttt{l + 1}; \\
       8 \quad & \texttt{\textcolor{gray}{// `var x = document.getElementById("b...'}}                                                                       \\
       9 \quad & \texttt{x} := \nil; \; \texttt{p} := \texttt{button}; \; \texttt{i} := \texttt{0}; \; \texttt{l} := \texttt{body}[\texttt{len}];            \\
      10 \quad & \kwwhile (\texttt{x} = \nil) \land (\texttt{i} \neq \texttt{l}) \;\{                                                                        \\
      11 \quad & \quad \texttt{i} := \texttt{i + 1}; \; \texttt{c} := \texttt{body}[\texttt{i}]; \; \texttt{n} := \texttt{c}[\texttt{id}];                   \\
      12 \quad & \quad \kwif \texttt{n} = \texttt{p} \kwblock{\texttt{x} := \texttt{c}} \kwelse \kwblock{\kwskip} \;\}                                       \\
      13 \quad & \texttt{\textcolor{gray}{// `if (x) x.innerHTML = x.getAttr...'}}                                                                           \\
      14 \quad & \kwif \texttt{x} \neq \nil \kwblock{\Output := \texttt{x}[\texttt{dataText}]} \kwelse \kwblock{\kwskip}                                     \\
    \end{aligned}
  \end{equation*}
  \captionsetup{type=lstlisting}
  \caption{
    DOM clobbering in $\modellanguagename$.
  }\label{listing:model:dom-script-gadgets}
\end{figure}

For this program we can demonstrate how the security conditions capture its exploitation. First, the input pair
\begin{equation*}
  \begin{aligned}
    &\Store_1[\Input_1 \mapsto 1, \Input_2 \mapsto 2, \Input_3 \mapsto 3, \Input_4 \mapsto 4]  \\
    &\Store_2[\Input_1 \mapsto 2, \Input_2 \mapsto 4, \Input_3 \mapsto 3, \Input_4 \mapsto 67] \\
  \end{aligned}
\end{equation*}
produces the traces (identical write observations omitted for brevity)
\begin{equation*}
  \begin{aligned}
    \trace_1 &= \multiset{\cdots, \observation^w_{[\![\texttt{el}]\!],1}}                                         \\
    \trace_2 &= \multiset{\cdots, \observation^w_{[\![\texttt{el}]\!],2}, \observation^r_{[\![\texttt{el}]\!],3}} \\
  \end{aligned}
\end{equation*}
given $\trustedmem = \emptyset$ we have $\sharedmem^u_i = \{ [\![\texttt{body}]\!], \cdots, [\![\texttt{el}]\!] \}$ and find
\begin{equation*}
  \begin{aligned}
                                 \trace_1 \cap \sharedmem^u_1 &\neq \trace_2 \cap \sharedmem^u_2                                                                      \\
    \multiset{\cdots, \observation^w_{[\![\texttt{el}]\!],1}} &\neq \multiset{\cdots, \observation^w_{[\![\texttt{el}]\!],2}, \observation^r_{[\![\texttt{el}]\!],3}} \\
  \end{aligned}
\end{equation*}
violating the security condition for a code reuse weakness.
If we consider the second execution, we can execute until $\observation^r_{[\![\texttt{el}]\!],2}$ (corresponding to $\texttt{c}[\texttt{id}]$, line 11) and continue execution with $v_1=3$ and $v_2=4$ until completion to observe different outputs, $\nil$ and $\texttt{67}$ resp., thus satisfying the security condition for a code reuse gadget.
Since we used an initial state of the weakness to demonstrate the gadget, and the modified observation is related to an observation in the delta between the two executions prior to it, namely $\observation^w_{[\![\texttt{el}]\!],2}$, the security condition for a code reuse vulnerability is also met.
Moreover, because $\trace^w_1 \cap \sharedmem^u_1 \neq \trace^w_2 \cap \sharedmem^u_2$ we know this is a \emph{stateful} vulnerability.

\begin{figure*}[]
  \begin{equation*}
    \begin{aligned}
      1 \quad & \texttt{shared} \gets \kwalloc_{\Context};                                                                                                   \\
      2 \quad & \texttt{x} := \texttt{shared}[\texttt{0}];                                                                                                   \\
      3 \quad & \kwif \texttt{x} = \texttt{1} \kwblock{ \texttt{y} := \texttt{shared}[\texttt{0}]; \; \Output := \texttt{y} } \kwelse \kwblock{\Output := 1} \\
    \end{aligned}
  \end{equation*}
  \captionsetup{type=lstlisting}
  \caption{
    While $\Output$ is guaranteed to be $\texttt{1}$ in practice, \Cref{definition:code-reuse-gadget} determines this program has a code reuse gadget because it can act on only the read observation on line 3.
  }\label{listing:model:definition-2-overapproximation}

  \begin{equation*}
    \begin{aligned}
      1 \quad & \texttt{shared} \gets \kwalloc_{\Context};                                                                \\
      2 \quad & \kwif \Input_1 = \texttt{1} \kwblock{\texttt{shared}[\texttt{0}] := \texttt{1}} \kwelse \kwblock{\kwskip} \\
      3 \quad & \texttt{x} := \texttt{shared}[\texttt{0}];                                                                \\
      4 \quad & \kwif \texttt{x} = \texttt{2} \kwblock{\Output := \texttt{67}} \kwelse \kwblock{\Output := \texttt{1}}    \\
    \end{aligned}
  \end{equation*}
  \captionsetup{type=lstlisting}
  \caption{
    While $\Output$ is guaranteed to be $\texttt{1}$ in practice, \Cref{definition:code-reuse-vulnerability} determines this program has a code reuse vulnerability because it has a weakness (line 2) and related gadget (line 3-4), but the combination of the then-branch on line 2 and the branch guard on line 4 prohibits a real vulnerability.
  }\label{listing:model:definition-3-overapproximation}

  \begin{equation*}
    \begin{aligned}
      1 \quad & \texttt{shared1} \gets \kwalloc_{\Context}; \; \texttt{shared1}[\Input_1] := \Input_2;                                                      \\
      2 \quad & \texttt{shared2} \gets \kwalloc_{\Context}; \; \texttt{shared2}[\texttt{0}] := \texttt{42}; \; \texttt{shared2}[\texttt{1}] := \texttt{67}; \\
      3 \quad & \Output := \texttt{shared2}[\Input_3]                                                                                                       \\
    \end{aligned}
  \end{equation*}
  \captionsetup{type=lstlisting}
  \caption{
    While this program has a \emph{stateless} vulnerability (line 2 and 3), the fact that it also has a \emph{stateful} \emph{weakness} (line 1) means this program is classified as having a \emph{stateful} code reuse vulnerability.
  }\label{listing:model:definition-stateless-stateful}

  \begin{equation*}
    \begin{aligned}
      1 \quad & \texttt{shared} \gets \kwalloc_{\Context};                                                                           \\
      2 \quad & \kwif \Input_1 = 1 \kwblock{\texttt{shared}[\texttt{1}] := \texttt{2}; \; \texttt{x} := \texttt{shared}[\texttt{1}]} \\
      3 \quad & \!\kwelse \kwblock{\texttt{x} := \texttt{shared}[\texttt{1}]; \; \texttt{shared}[\texttt{1}] := \texttt{2}}          \\
      4 \quad & \Output := \texttt{x}                                                                                                \\
    \end{aligned}
  \end{equation*}
  \captionsetup{type=lstlisting}
  \caption{
    This program enables an attacker to influence the order in which shared state is accessed without the ability to change the locations being accessed.
    As such, for two executions starting from $\Store_1[\Input_1 \mapsto 0]$ and $\Store_2[\Input_1 \mapsto 1]$, the multiset of observations is identical but the sequence would not be.
  }\label{listing:model:observation-order}
\end{figure*}

\begin{table*}[]
  \centering
  \caption{
    Overview of (minimal) code reuse gadgets.
    Each \textbf{gadget} is written in a JavaScript style and should be considered to have the prefix \texttt{x = obj[u]} for reading from the prototype using untrusted \texttt{u}.
    Every \textbf{Path} starts from \texttt{Data flow}, omitted for brevity.
  }\label{table:code-reuse-gadgets}
  \begin{tabular}{ r l l l }
    \textbf{\#} & \textbf{Gadget}                                 & \textbf{Framework Path}                                                                                                   & \textbf{Type} \\
      \midrule
     1          & \texttt{sink(x)}                                & $\ldots\!\to$\texttt{Sink}                                                                                                & Simple        \\
     2          & \texttt{x() --> sink()}                         & $\ldots\!\to$\texttt{Call flow}$\to$\texttt{Sink}                                                                         & Simple        \\
     3          & \texttt{if(x)\{sink()\}}                        & $\ldots\!\to$\texttt{Control flow}$\to$\texttt{Sink}                                                                      & Simple        \\
     4          & \texttt{if(x)\{g()\} --> sink()}                & $\ldots\!\to$\texttt{Control flow}$\to$\texttt{Call flow}$\to$\texttt{Sink}                                               & Simple        \\
      \midrule
     5          & \texttt{if(x)\{y=a\}; sink(y)}                  & $\ldots\!\to$\texttt{Control flow}$\to$\texttt{Data flow}$\to$\texttt{Sink}                                               & Vertical      \\
     6          & \texttt{if(x)\{f=g\}; f() --> sink()}           & $\ldots\!\to$\texttt{Control flow}$\to$\texttt{Data flow}$\to$\texttt{Call flow}$\to$\texttt{Sink}                        & Vertical      \\
     7          & \texttt{if(x)\{f=g\}; y=f(); sink(y)}           & $\ldots\!\to$\texttt{Control flow}$\to$\texttt{Data flow}$\to$\texttt{Call flow}$\to$\texttt{Data flow}$\to$\texttt{Sink} & Vertical      \\
     8          & \texttt{if(x)\{y=f()\}; sink(y)}                & $\ldots\!\to$\texttt{Control flow}$\to$\texttt{Call flow}$\to$\texttt{Data flow}$\to$\texttt{Sink}                        & Vertical      \\
     9          & \texttt{if(x)\{g=f()\}; g() --> sink()}         & $\ldots\!\to$\texttt{Control flow}$\to$\texttt{Call flow}$\to$\texttt{Data flow}$\to$\texttt{Call flow}$\to$\texttt{Sink} & Vertical      \\
    10          & \texttt{y=x(); sink(y)}                         & $\ldots\!\to$\texttt{Call flow}$\to$\texttt{Data flow}$\to$\texttt{Sink}                                                  & Vertical      \\
    11          & \texttt{y=x(); if(y)\{sink()\}}                 & $\ldots\!\to$\texttt{Call flow}$\to$\texttt{Data flow}$\to$\texttt{Control flow}$\to$\texttt{Sink}                        & Vertical      \\
    12          & \texttt{y=x(); if(y)\{z=a\}; sink(z)}           & $\ldots\!\to$\texttt{Call flow}$\to$\texttt{Data flow}$\to$\texttt{Control flow}$\to$\texttt{Data flow}$\to$\texttt{Sink} & Vertical      \\
    13          & \texttt{y=x(); if(y)\{f() --> sink()\}}         & $\ldots\!\to$\texttt{Call flow}$\to$\texttt{Data flow}$\to$\texttt{Control flow}$\to$\texttt{Call flow}$\to$\texttt{Sink} & Vertical      \\
      \midrule
    14          & \texttt{y=obj[x]; sink(y)}                      & $\ldots\!\to$Fetch$\to\!\ldots$                                                                                           & Horizontal    \\
    15          & \texttt{x() --> \{y=obj[q]; sink(y)\}}          & $\ldots\!\to$\texttt{Call flow}$\to$Fetch$\to\!\ldots$                                                                    & Horizontal    \\
    16          & \texttt{if(x)\{y=obj[q]; sink(y)\}}             & $\ldots\!\to$\texttt{Control flow}$\to$Fetch$\to\!\ldots$                                                                 & Horizontal    \\
    17          & \texttt{if(x)\{g()\} --> \{y=obj[q]; sink(y)\}} & $\ldots\!\to$\texttt{Control flow}$\to$\texttt{Call flow}$\to$Fetch$\to\!\ldots$                                          & Horizontal    \\
  \end{tabular}
\end{table*}

\begin{table*}[]
  \centering
  \caption{
    Overview of all defenses mapped to their corresponding Strategy and Approach from \Cref{table:defenses-1}, whether it Targets the weakness or gadget, and the framework (\Cref{figure:framework}) edge it cuts.
    Defenses proposed by this work are \textit{italicized}.
  }\label{table:defenses-3}
  \begin{tabular}{ r l l l l }
                                        & \textbf{Strategy} & \textbf{Approach}       & \textbf{Target}   & \textbf{Framework Edge}               \\
      \midrule
    Type casting                        & Mediate           & Allow/Blocklist         & Weakness          & $\texttt{Data}\to\texttt{Fetch}$      \\
    Look-ahead object parsing           & Mediate           & Allow/Blocklist         & Weakness          & $\texttt{Data}\to\texttt{Fetch}$      \\
    Debloating                          & Minimize          & Remove                  & Weakness          & $\texttt{Data}\to\texttt{Fetch}$      \\
    \textit{Use static methods}         & \textit{Minimize} & \textit{Avoid}          & \textit{Weakness} & $\texttt{Data}\to\texttt{Fetch}$      \\
      \midrule
    Property filtering                  & Mediate           & Allow/Blocklist         & Weakness          & $\texttt{Data}\to\texttt{Store}$      \\
    Object.freeze()                     & Constrain         & Immutability            & Weakness          & $\texttt{Data}\to\texttt{Store}$      \\
    Object.seal()                       & Constrain         & Extensibility           & Weakness          & $\texttt{Data}\to\texttt{Store}$      \\
    Object.preventExtension()           & Constrain         & Extensibility           & Weakness          & $\texttt{Data}\to\texttt{Store}$      \\
    Symbol.proto                        & Mediate           & Stratification          & Weakness          & $\texttt{Data}\to\texttt{Store}$      \\
    Object.hasOwn()                     & Mediate           & Validation              & Gadget            & $\texttt{Fetch}\to\texttt{Data flow}$ \\
    new Map()                           & Mediate           & Abstraction             & Gadget            & $\texttt{Fetch}\to\texttt{Data flow}$ \\ %
    Object.create(null)                 & Mediate           & Abstraction             & Gadget            & $\texttt{Fetch}\to\texttt{Data flow}$ \\ %
    \textit{Use bare/static functions}  & \textit{Minimize} & \textit{Avoid}          & \textit{Weakness} & $\texttt{Data}\to\texttt{Fetch}$      \\
      \midrule
    Property filtering                  & Mediate           & Allow/Blocklist         & Weakness          & $\texttt{Data}\to\texttt{Store}$      \\
    \textit{Stratify dunder properties} & \textit{Mediate}  & \textit{Stratification} & \textit{Weakness} & $\texttt{Data}\to\texttt{Store}$      \\
      \midrule
    Disable named properties            & Minimize          & Avoid                   & Weakness          & $\texttt{Data}\to\texttt{Store}$      \\
    Remove (bad) name/id                & Mediate           & Allow/Blocklist         & Weakness          & $\texttt{Data}\to\texttt{Store}$      \\
    Prefix name/id                      & Mediate           & Namespace               & Weakness          & $\texttt{Store}\to\texttt{Fetch}$     \\
    Object.defineProperty()             & Constrain         & Immutability            & Weakness          & $\texttt{Data}\to\texttt{Store}$      \\
    Unconditional globals               & Constrain         & Immutability            & Weakness          & $\texttt{Data}\to\texttt{Store}$      \\
    Type checking                       & Mediate           & Validation              & Gadget            & $\texttt{Fetch}\to\texttt{Data flow}$ \\
    \textit{Custom API}                 & \textit{Mediate}  & \textit{Abstraction}    & \textit{Gadget}   & $\texttt{Fetch}\to\texttt{Data flow}$ \\
     \midrule
    Input validation                    & Mediate           & Validation              & Gadget            & $\texttt{Data flow}\to\texttt{Sink}$  \\
    \textit{No DOM state}               & \textit{Minimize} & \textit{Avoid}          & \textit{Gadget}   & $\texttt{Fetch}\to\texttt{Data flow}$ \\
  \end{tabular}
\end{table*}

\end{document}